\documentclass[aps,prb,reprint,amsmath,amssymb,groupedaddress]{revtex4-2}
\usepackage{bm}
\usepackage{graphicx}
\usepackage{dcolumn}

\begin{document}

\title{
Anisotropic Behavior of the Thermoelectric Power and the Thermal Conductivity in a Unidirectional Lateral Superlattice: 
A Typical Anisotropic System Exhibiting Two Distinct Nernst Coefficients}


\author{Akira Endo}
\email[]{akrendo@issp.u-tokyo.ac.jp}
\affiliation{The Institute for Solid State Physics, The University of Tokyo, 5-1-5 Kashiwanoha, Kashiwa, Chiba 277-8581, Japan}

\author{Shingo Katsumoto}
\affiliation{The Institute for Solid State Physics, The University of Tokyo, 5-1-5 Kashiwanoha, Kashiwa, Chiba 277-8581, Japan}

\author{Yasuhiro Iye}
\affiliation{The Institute for Solid State Physics, The University of Tokyo, 5-1-5 Kashiwanoha, Kashiwa, Chiba 277-8581, Japan}


\date{\today}

\begin{abstract}
We have calculated the thermoelectric conductivity tensor $\varepsilon_{ij}$ and the thermal conductivity tensor $\lambda_{ij}$ of a unidirectional lateral superlattice (ULSL) ($i,j = x,y$, with the $x$-axis aligned to the principal axis of the ULSL), 
based on the asymptotic analytic formulas of the electrical conductivity tensor $\sigma_{ij}$ in the literature valid at low magnetic fields where large numbers of Landau levels are occupied. With the resulting analytic expressions, we clarify the conditions for the Mott formula (Wiedemann-Franz law) to be applicable  with high precision to $\varepsilon_{ij}$ ($\lambda_{ij}$). We further present plots of the commensurability oscillations $\delta\varepsilon_{ij}$, $\delta\lambda_{ij}$, $\delta\kappa_{ij}$, and $\delta S_{ij}$ in $\varepsilon_{ij}$, $\lambda_{ij}$, (an alternative, more standard definition of) the thermal conductivity tensor $\kappa_{ij}$, and the thermopower tensor $S_{ij}$, calculated using typical parameters for a ULSL fabricated from a GaAs/AlGaAs two-dimensional electron gas (2DEG). Notable features of the $\delta S_{ij}$ are (i) anisotropic behavior ($\delta S_{xx} \ne \delta S_{yy}$) and (ii) the dominance of the $xy$ component over the other components ($|\delta S_{xy}| \gg |\delta S_{yx}|, |\delta S_{xx}|, |\delta S_{yy}|$). The latter clearly indicates that the two Nernst coefficients, $S_{xy}$ and $S_{yx}$, can be totally different from each other in an anisotropic system. Both (i) and (ii) are at variance with the previous theory and are attributable to the inclusion of a damping factor due to the small-angle scattering characteristic of GaAs/AlGaAs 2DEGs, which have not been taken into consideration in $\delta S_{ij}$ thus far.
\end{abstract}

\maketitle

\newpage

\section{Introduction}
A broken symmetry often plays a pivotal role in condensed matter physics. In the present paper, we consider a system possessing arguably the simplest form of a broken rotational symmetry, namely, a unidirectional lateral superlattice (ULSL), in which twofold in-plane anisotropy is artificially introduced into an otherwise isotropic two-dimensional electron gas (2DEG) by imposing a unidirectional periodic potential modulation. We focus on the thermoelectric and thermal transport coefficients, which are currently  the subject of growing interest as an effective tool to probe exotic electronic states difficult to be accessed by more widely studied electrical transport coefficients \cite{Yang09,Barlas12,Venkatachalam12,Banerjee17,Srivastav22}. Main interest in the present study is to see how the anisotropy manifests itself in the thermoelectric/thermal transport coefficients when placed in a magnetic field and whether or not qualitative difference is found in the appearance of the anisotropy compared with that in the electrical transport coefficients.

One of the best-known magnetotransport phenomena in a ULSL is the commensurability oscillations (COs) resulting from the commensurability between the period $a$ of the unidirectional potential modulation and the cyclotron radius $R_\mathrm{c} = \hbar k_\mathrm{F} / (e|B|)$, where $k_\mathrm{F} = \sqrt{2 \pi n_e}$ is the Fermi wavenumber with $n_e$ the areal density of the 2DEG\@. Here and throughout the paper, we set the 2DEG plane as the $x$-$y$ plane, the principal axis of the modulation as the $x$-axis, and the direction of the magnetic field $B$ as the $z$-axis. Following the experimental discovery in the resistivities $\rho_{xx}$ and $\rho_{yy}$ (Weiss oscillations) \cite{Weiss89},  various electrical, thermal, and thermodynamic quantities in a ULSL were theoretically predicted \cite{Peeters92} to oscillate with a magnetic field $B$. Experimentally, COs have further been observed in the capacitance \cite{Weiss89C}, thermopower \cite{Taboryski95,Nogaret02,Endo11IC,KoikeMT,KoikeEP2DS20}, electrical conductivity \cite{KajiokaCO13}, and recently also in the Hall resistance \cite{Endo21HallCO}.

In Ref.\ \onlinecite{Peeters92}, the calculated electrical, thermoelectric, and thermal transport coefficients were mainly presented in the form of the summation of the terms containing the (associated) Laguerre polynomials over the Landau index $N$. Thus, numerical calculations are required to perceive the behaviors of the transport coefficients. For the conductivities $\sigma_{xx}$ and $\sigma_{yy}$, however, analytic formulas were also provided \cite{Peeters92}, which allows us to readily grasp the dependence of the amplitude and the phase of the COs on the magnetic field $B$, the temperature $T$, and the parameters of the ULSL sample. The analytic formulas were obtained by replacing the (associated) Laguerre polynomials by their asymptotic expressions at large $N$. Generally, many Landau levels are occupied in a typical ULSL ($n_e \gtrsim 10^{15}$ m$^{-2}$, $a \gtrsim 100$ nm)  in the magnetic-field range COs are observed ($B \lesssim 1$ T), validating the use of the asymptotic expressions. Following the same strategy, the present authors deduced an approximate analytic expression also for $\sigma_{yx}$ in a recent publication \cite{Endo21HallCO}. 

One of the purposes of the present paper is to present similar analytic formulas for the thermoelectric power and the thermal conductivity. The formulas are obtained by straightforward calculations of the first- and the second-order moments performed on the analytic expressions for  $\sigma_{xx}$, $\sigma_{yy}$, and $\sigma_{yx}$ just mentioned above, with a slight modification to include an additional damping factor accounting for the effect of the small-angle scattering that impede the completion of a cycle of the cyclotron orbit \cite{Mirlin98,Endo00e}. The thermoelectric power and the thermal conductivity are often deduced approximately from the electrical conductivity via the Mott formula and the Wiedemann-Franz law, respectively at low temperatures. We examine the conditions for these approximations to be accurate using the obtained analytic expressions. 

We present examples of COs calculated by these analytic formulas at temperatures ranging from 0.02 K to 10 K, plugging in the parameters for a typical ULSL sample fabricated from a conventional 2DEG residing at the GaAs/AlGaAs heterointerface. The plots clearly reveal the deviation from the Mott formula or the Wiedemann-Franz law with increasing temperatures. Important aspects worth highlighting  in the COs of the thermopower tensor are the anisotropic behavior of the diagonal (Seebeck) components and the far dominance of the $xy$ component, one of the off-diagonal (Nernst) component, over the other Seebeck and Nernst components. These behaviors are at variance with the previous theory \cite{Peeters92}. The disagreement mainly results from the inclusion of the additional damping factor mentioned above \cite{Mirlin98,Endo00e}, which was shown to be indispensable for achieving good agreement between theoretical and experimental COs in the resistivity \cite{Endo00e} and the Hall resistance \cite{Endo21HallCO}.
We will also comment on the difficulties, caused by the tilting of the temperature gradient in the magnetic field \cite{Endo19}, in the interpretation of the experimentally observed COs in the thermopower.

\section{Transport Coefficients}
First we outline the general properties of the thermoelectric and thermal transport coefficients to be calculated below. The electrical conductivity tensor $\hat{\sigma}$, the thermoelectric conductivity tensor $\hat{\varepsilon}$, the Peltier tensor $\hat{\pi}$, and the thermal conductivity tensor $\hat{\lambda}$ are defined as the coefficients relating the electrical current density $\boldsymbol{j}$ and the thermal current density $\boldsymbol{j_Q}$ to the electric field $\boldsymbol{E}$ and the temperature gradient $-\boldsymbol{\nabla} T$ in the linear response regime \cite{Fletcher99,ZimanEP60,AshcroftMermin76,Pottier10}\footnote{The notations of the coefficients here follows those of Ref.\ \onlinecite{Fletcher99}. $\sigma$, $\varepsilon$, $\pi$, and $\lambda$ are denoted as $L_{EE}$, $-L_{ET}$, $L_{TE}$, and $-L_{TT}$ in Ref.\ \onlinecite{ZimanEP60} and $e^2L_{11}/T$, $-eL_{12}/T$, $-eL_{21}/T$, and $L_{22}/T$ in Ref.\ \onlinecite{Pottier10}}
\begin{equation}
\boldsymbol{j} = \hat{\sigma} \boldsymbol{E}-\hat{\varepsilon} \boldsymbol{\nabla} T \label{jET}
\end{equation}
\begin{equation}
\boldsymbol{j_Q} = \hat{\pi} \boldsymbol{E}-\hat{\lambda} \boldsymbol{\nabla} T, \label{jQET}
\end{equation}
where we used the `hat' $\hat{X}$ to denote the $2\times 2$ tensor containing the components $X_{ij}$ ($i, j = x, y$).
Alternative linear response equations relating $\boldsymbol{E}$ and $\boldsymbol{j_Q}$ to $\boldsymbol{j}$ and $-\boldsymbol{\nabla} T$ \cite{ZimanEP60},
\begin{equation}
\boldsymbol{E} = \hat{\rho} \boldsymbol{j}+\hat{S} \boldsymbol{\nabla} T \label{EjT}
\end{equation}
\begin{equation}
\boldsymbol{j_Q} = \hat{\Pi} \boldsymbol{j}-\hat{\kappa} \boldsymbol{\nabla} T, \label{jQjT}
\end{equation}
define the resistivity tensor $\hat{\rho}$, the thermopower tensor $\hat{S}$, and (more standard definitions of) the Peltier and the thermal conductivity tensors, $\hat{\Pi}$ and $\hat{\kappa}$. By substituting Eq.\ (\ref{EjT}) to Eqs.\ (\ref{jET}) and (\ref{jQET}), one can readily find the relations between the coefficients,
\begin{equation}
\hat{\rho} = \hat{\sigma}^{-1}  \label{rhosgm}
\end{equation}
\begin{equation}
\hat{S} = \hat{\sigma}^{-1} \hat{\varepsilon} = \hat{\rho} \hat{\varepsilon} \label{Srhoeps}
\end{equation}
\begin{equation}
\hat{\Pi} = \hat{\pi} \hat{\rho} \label{Pipi}
\end{equation}
\begin{equation}
\hat{\kappa} = \hat{\lambda} - \hat{\pi} \hat{S} = \hat{\lambda} - \hat{\pi} \hat{\rho} \hat{\varepsilon}. \label{kappalambda}
\end{equation}
According to the Kelvin-Onsager relations \cite{OnsagerI31,OnsagerII31,ZimanEP60,Pottier10}, we further have
\begin{subequations}
\begin{eqnarray}
\hat{\sigma}(B) & = & ^t\! \hat{\sigma}(-B) \label{Onsagersgm} \\
\hat{\lambda}(B) & = & ^t\! \hat{\lambda}(-B) \label{Onsagerlmd} \\
\hat{\pi}(B) & = & T\  ^t\! \hat{\varepsilon}(-B), \label{Onsagerpi} 
\end{eqnarray}
\end{subequations}
which, combined with Eqs.\ (\ref{rhosgm}), (\ref{Srhoeps}), and (\ref{Pipi}), lead also to the relations
\begin{subequations}
\begin{eqnarray}
\hat{\rho}(B) & = & ^t\! \hat{\rho}(-B) \label{Onsagerrho} \\
\hat{\kappa}(B) & = & ^t\! \hat{\kappa}(-B) \label{Onsagerkappa} \\
\hat{\Pi}(B) & = & T\  ^t\! \hat{S}(-B). \label{OnsagerPi} 
\end{eqnarray}
\end{subequations}
In general, the thermoelectric conductivity $\hat{\varepsilon}$ is composed of two components: the diffusion contribution and the phonon-drag contribution \cite{Fletcher99,GoldsmidBk10}. In the present study, we limit ourselves to the former contribution. The latter contribution can become the dominant contribution in GaAs/AlGaAs 2DEGs in the standard experimental setup using an external heater to introduce the temperature gradient not only in the 2DEG but also in the lattice hosting the 2DEG \cite{Fletcher99,Fletcher86}. We can, however, eliminate the phonon-drag contribution by confining ourselves to very low temperatures ($\lesssim 0.2$ K) where the electron-phonon interaction becomes negligibly small \cite{KajiokaCO13,Liu18} or by employing the current heating technique \cite{Maximov04,Fujita10E}, which introduce the temperature gradient only into the 2DEG without affecting the lattice temperature.
Similarly, we only consider the thermal conductivity due to the electrons and neglect the phonon contribution. The diffusion thermoelectric conductivity and the electronic thermal conductivity are related to the electrical conductivity at $T = 0$ as \cite{Peeters92,Fletcher99,ZimanEP60,AshcroftMermin76,Jonson84}
\begin{equation}
\varepsilon_{ij}=-\frac{1}{eT}\int_{-\infty}^{\infty}dE\left[-f^\prime\left(E\right)\right]\left(E-E_\mathrm{F}\right)\sigma_{ij,{T=0}}\left(E\right) \label{epsilondef}
\end{equation}
and
\begin{equation}
\lambda_{ij}=\frac{1}{e^2T}\int_{-\infty}^{\infty}dE\left[-f^\prime\left(E\right)\right]\left(E-E_\mathrm{F}\right)^2\sigma_{ij,{T=0}}\left(E\right), \label{lambdadef}
\end{equation}
respectively. 
Here $f^\prime(E)$ is the energy derivative of the Fermi-Dirac distribution function $f(E) = \{1+\exp[(E-E_\mathrm{F})/k_\mathrm{B}T]\}^{-1}$, with $E_\mathrm{F}$ the Fermi energy and $k_\mathrm{B}$ the Boltzmann constant. By expanding $\sigma_{ij,{T=0}}(E)$ around $E_\mathrm{F}$ (Sommerfeld expansion),
\begin{equation}
\sigma_{ij,{T=0}}(E) = \sigma_{i j,{T=0}}(E_\mathrm{F}) + \left.\frac{\partial\sigma_{i j,{T=0}}}{\partial E}\right|_{E_\mathrm{F}} \! \! \! \! \! \! \! (E - E_\mathrm{F}) + O[(E - E_\mathrm{F})^2], \label{sgmexpnd}
\end{equation}
and retaining only the lowest order in $E - E_\mathrm{F}$, we arrive at the Mott formula \cite{Jonson84,ZimanEP60,AshcroftMermin76}
\begin{equation}
\varepsilon_{i j}^\mathrm{M} =-L_0eT\left.\frac{\partial\sigma_{i j,{T=0}}(E)}{\partial E}\right|_{E_\mathrm{F}} \label{MottRel}
\end{equation}
and the Wiedemann-Franz law \cite{ZimanEP60,AshcroftMermin76}
\begin{equation}
\lambda_{ij}^\mathrm{W} = L_0T \sigma_{i j,{T=0}}(E_\mathrm{F}) \label{WFlaw}
\end{equation}
as low-temperature approximations, where $L_0 = \pi^2 {k_\mathrm{B}}^2/(3e^2) = 2.44\times10^{-8}$ V$^2$/K$^2$ is the Lorenz number. We will discuss the temperature range for Eqs.\ (\ref{MottRel}) and (\ref{WFlaw}) to be valid for COs.

The Kelvin-Onsager relations Eqs.\ (\ref{Onsagersgm}), (\ref{Onsagerlmd}), (\ref{Onsagerrho}), and (\ref{Onsagerkappa}), along with the antisymmetry $X_{ij}(-B) = -X_{ij}(B)$  ($i \ne j$) of the off-diagonal components under magnetic-field reversal, yields the relations $\sigma_{xy} = -\sigma_{yx}$, $\lambda_{xy} = -\lambda_{yx}$,  $\rho_{xy} = -\rho_{yx}$, and $\kappa_{xy} = -\kappa_{yx}$. In addition, the diffusion thermoelectric conductivity also possesses similar relation $\varepsilon_{xy} = -\varepsilon_{yx}$ owing to Eq.\ (\ref{epsilondef}). 
For $\hat{\sigma}$, $\hat{\varepsilon}$, $\hat{\lambda}$, and $\hat{\kappa}$, therefore, we only need to calculate $xx$, $yy$, and $yx$ components.
For the thermopower tensor $\hat{S}$, however, the Kelvin-Onsager relation, Eq.\ (\ref{OnsagerPi}), does not place any constraints on the relation between $S_{xy}$ and $S_{yx}$.
Since $S_{xy} = -\rho_{xx} \varepsilon_{yx}+\rho_{xy} \varepsilon_{yy}$ and $S_{yx} = -\rho_{xy} \varepsilon_{xx} + \rho_{yy} \varepsilon_{yx}$, the relation $S_{xy} = -S_{yx}$ holds for isotropic systems with $\rho_{xx} = \rho_{yy}$ and $\varepsilon_{xx} = \varepsilon_{yy}$. In anisotropic systems, however, the relation is not generally valid. A ULSL provides an archetypal example for how anisotropy affects the relation between the two Nernst components, $S_{xy}$ and $S_{yx}$. We thus need to calculate both $S_{xy}$ and $S_{yx}$ separately, in addition to $S_{xx}$ and $S_{yy}$. As we will see, the calculations reveal considerable difference between the two Nernst components.

\section{Analytic Formulas}
\subsection{Electrical conductivity and resistivity}
\begin{figure}[t]
\includegraphics[width=8.6cm,clip]{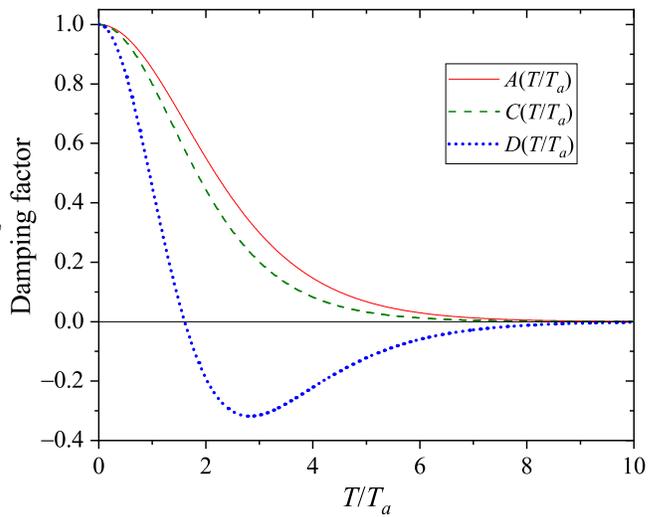}%
\caption{(Color online) Plots of thermal damping factors $A(T/T_a)$ (solid red line), $B(T/T_a)$ (dashed dark-green line), and $C(T/T_a)$ (dotted blue line) given by Eqs.\ (\ref{EqA}), (\ref{EqC}), and (\ref{EqD}) respectively. \label{ACD}}
\end{figure}
\begin{figure}[t]
\includegraphics[width=8.6cm,clip]{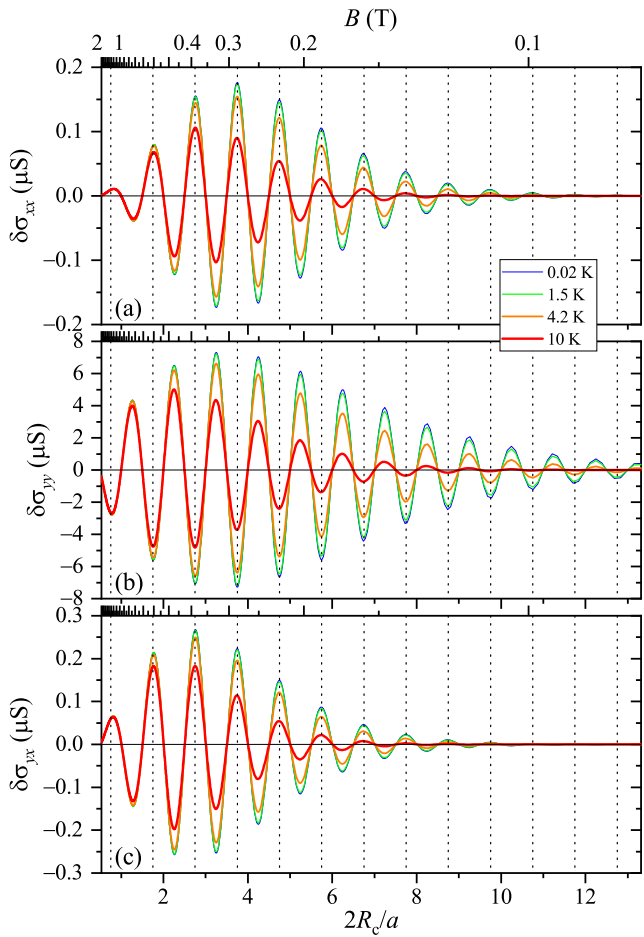}%
\caption{(Color online) Commensurability oscillations in the electrical conductivity vs.\ $2 R_\mathrm{c} / a$ calculated for various temperatures by picking out the oscillatory parts from Eq.\ (\ref{sgmall}) and using typical ULSL parameters: $V_0 = 0.26$ meV, $a = 200$ nm, $\mu_\mathrm{b} = 5.3$ m$^2$/(Vs), $\mu_\mathrm{c} = \mu_\mathrm{h} = \mu_\mathrm{b}/2$, $n_e = 4.2\times10^{15}$ m$^{-2}$, and $\mu = 72$ m$^2$/(Vs). The top axis shows $B$. (a) $\delta \sigma_{xx}$, (b) $\delta \sigma_{yy}$, and (c) $\delta \sigma_{yx}$.  Vertical dotted lines indicate the locations of the flat-band conditions Eq.\ (\ref{flatband}). \label{Gsgm}}
\end{figure}
\begin{figure}[t]
\includegraphics[width=8.6cm,clip]{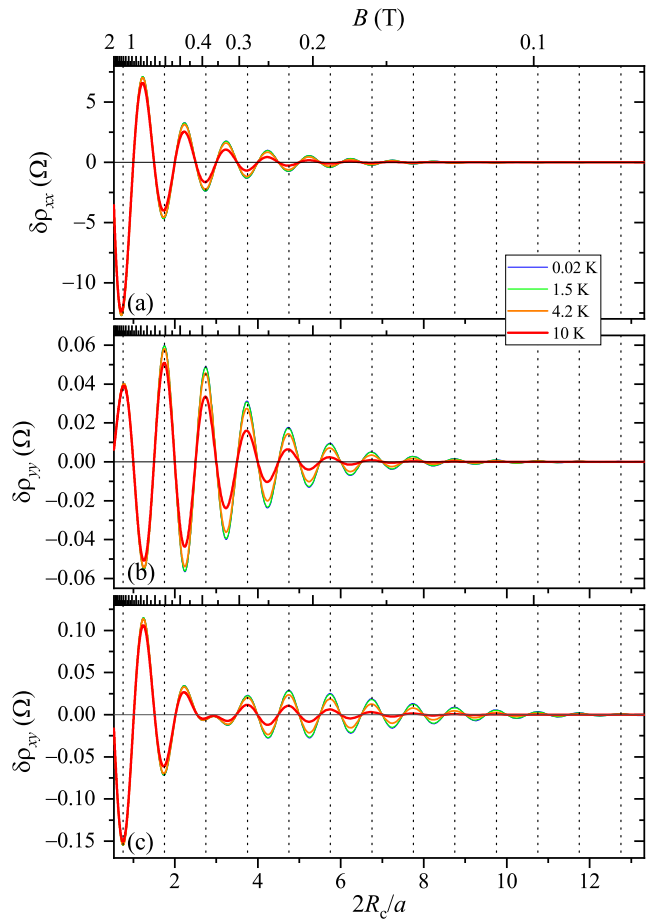}%
\caption{(Color online) Commensurability oscillations in the resistivity vs.\ $2 R_\mathrm{c} / a$ calculated for various temperatures by Eq.\ (\ref{drhoall}) using the same ULSL parameters as in Fig.\ \ref{Gsgm}. The top axis shows $B$. (a) $\delta \rho_{xx}$, (b) $\delta \rho_{yy}$, and (c) $\delta \rho_{xy}$. Vertical dotted lines indicate the locations of the flat-band conditions Eq.\ (\ref{flatband}). \label{Grho}}
\end{figure}

We start by reviewing the well-documented analytic asymptotic expressions for the electrical conductivity and resistivity tensors. We assume a sinusoidal modulation 
$V(x) = V_0 \cos(2 \pi x / a)$ with small amplitude $V_0 / E_\mathrm{F} \ll 1$.
It is well known that the COs in the electrical conductivity arise from two separate mechanisms: the band contribution and the collisional contribution \cite{Peeters92}. The former originates from the oscillations of the $y$-component of the group velocity (band dispersion) with $B$,  and is explained also by a semiclassical picture invoking the guiding center drift of the cyclotron motion in the $y$ direction \cite{Beenakker89}. The band contribution is thus contained only in $\sigma_{yy}$ and accounts for the major part in its COs, $\delta \sigma_{yy}$. Here and in what follows, we denote the oscillatory part (COs) contained in the coefficient $X_{ij}$ by $\delta X_{ij}$. 
The asymptotic expression for the band contribution in the electrical conductivity is given as \cite{Peeters92}
\begin{equation}
\sigma_{yy}^\mathrm{band} = \frac{\sigma_0 V_0^2}{E_\mathrm{F} \hbar \omega_\mathrm{c} a k_\mathrm{F}}\left[1 + A\left( \frac{\pi}{\mu_\mathrm{b} B} \right) A\left( \frac{T}{T_a} \right) \sin{r_\mathrm{c}} \right], \label{sgmyyband}
\end{equation}
where $\sigma_0 = e n_e \mu$ is the electrical conductivity at $B = 0$ with $\mu$ the mobility, $\omega_\mathrm{c} = e|B|/m^*$ is the cyclotron angular frequency with $m^*$ the effective mass ($m^* =0.067 m_e$  for GaAs with $m_e$ the electron mass), and $r_\mathrm{c} \equiv 2 \pi (2 R_\mathrm{c} / a)$ is the reduced cyclotron radius. The temperature dependence of the CO amplitude is dictated by the thermal damping factor $A(T/T_a)$ (shown in Fig.\ \ref{ACD}), with $T_a \equiv \hbar \omega_\mathrm{c} a k_\mathrm{F} / (4 \pi^2 k_\mathrm{B})$ representing the characteristic temperature and
\begin{equation}
A\left(x\right)\equiv\frac{x}{\sinh{x}}=1-\frac{x^2}{6}+O\left(x^4\right). \label{EqA}
\end{equation}
Although absent in the original theory \cite{Peeters92}, an additional damping factor $A(\pi/\mu_\mathrm{b}B)$ is included to account for the effect of the small-angle scattering that deflects the electrons out of the cyclotron trajectory before completing a cycle. The damping factor was theoretically deduced using the Boltzmann equation formalism in Ref.\ \onlinecite{Mirlin98}. The present authors showed that an excellent agreement with experimental COs is obtained with the value of the parameter $\mu_\mathrm{b}$ close to the quantum mobility $\mu_\mathrm{q}$ deduced from the damping of the Shubnikov-de Haas (SdH) oscillations \cite{Endo00e}.
The collisional contribution, on the other hand, is caused by the oscillations in the density of states (DOS), analogous to SdH oscillations, and thus isotropically affects the electrical conductivity. The asymptotic expression \cite{Peeters92} is written as
\begin{equation}
\sigma_{xx}^\mathrm{col} = \frac{3 \sigma_0 V_0^2 a k_\mathrm{F}}{8 \pi^2 E_\mathrm{F} \hbar \omega_\mathrm{c} \mu^2 B^2}\left[1 - A\left( \frac{\pi}{\mu_\mathrm{c} B} \right) A\left( \frac{T}{T_a} \right) \sin{r_\mathrm{c}} \right]. \label{sgmxxcol}
\end{equation}
Here, we simply assumed that the effect of small-angle scattering can be accounted for by a damping factor $A(\pi/\mu_\mathrm{c}B)$ similar to that in  the band contribution, although the theory for the collisional contribution that takes the small-angle scattering into account is still missing to the best of our knowledge.  Note, however, that, without the damping factor, the amplitude of COs in $\sigma_{xx}$ and $\sigma_{yx}$ increases with decreasing magnetic field and diverges with $B \rightarrow 0$, which is in clear disagreement with the experiments \cite{KajiokaCO13,Endo21HallCO}. We allowed the parameter describing the damping, $\mu_\mathrm{c}$, to differ from $\mu_\mathrm{b}$ in Eq.\ (\ref{sgmyyband}), admitting the possibility that the effect of the small-angle scattering on the COs varies for a distinct mechanism. 
The oscillatory parts, $\delta \sigma_{yy}^\mathrm{band}$ and $\delta \sigma_{xx}^\mathrm{col}$, are given by the second terms in Eqs.\ (\ref{sgmyyband}) and (\ref{sgmxxcol}), respectively. We can readily see that $\delta \sigma_{yy}^\mathrm{band}$ and $\delta \sigma_{xx}^\mathrm{col}$ oscillate with opposite phases, with the former (latter) taking minima (maxima) at the flat-band conditions,
\begin{equation}
\frac{r_\mathrm{c}}{2\pi} = \frac{2 R_\mathrm{c}}{a} = n - \frac{1}{4}, \hspace{5mm} (n = 1, 2, 3,...), \label{flatband}
\end{equation}
and that $|\delta \sigma_{yy}^\mathrm{band}| \gg |\delta \sigma_{xx}^\mathrm{col}|$ noting that generally $\mu B \gg 1$ in the magnetic-field range COs are observed. 

The Hall conductivity also contains COs originating from the collisional contribution. However, asymptotic expressions were not presented in the original theory \cite{Peeters92}. From the formula given as the summation over Landau index $N$, Eq.\ (28) in Ref.\ \onlinecite{Peeters92}, the present authors have recently deduced approximate analytic formulas presented as the sum of the following three constituents \cite{Endo21HallCO},
\begin{widetext}
\begin{subequations} \label{sgmyxall}
\begin{equation}
\sigma_{yx}^{\left(1\right)} = \frac{3}{2}\nu\frac{e^2}{h}{\lambda_\mathrm{c}}^2 \left[1- A\left(\frac{\pi}{\mu_\mathrm{h} B}\right)A\left(\frac{T}{T_a}\right)\sin{\left(r_\mathrm{c}+\delta_\mathrm{F}\right)}\right], \label{sgmyx1}
\end{equation}
\begin{equation}
\sigma_{yx}^{\left(21\right)} = -\nu\frac{e^2}{h}{\lambda_\mathrm{c}}^2\frac{ak_\mathrm{F}}{\pi}\sin{\left(\frac{\pi}{{ak}_\mathrm{F}}\right)}\left[\cos{\left(\frac{\delta_\mathrm{F}}{2}+\frac{\pi}{{ak}_\mathrm{F}}\right)}+A\left(\frac{\pi}{\mu_\mathrm{h} B}\right)A\left(\frac{T}{T_a}\right)\sin{\left(r_\mathrm{c}+\frac{\delta_\mathrm{F}}{2}-\frac{\pi}{{ak}_\mathrm{F}}\right)}\right], \label{sgmyx21}
\end{equation}
and
\begin{equation}
\sigma_{yx}^{\left(22\right)} = -\mathrm{sgn}(B)\frac{e^2}{h}{\lambda_\mathrm{c}}^2\frac{ak_\mathrm{F}}{\pi}\cos{\left(\frac{\pi}{{ak}_\mathrm{F}}\right)}\left[ \sin{\left(\frac{\delta_\mathrm{F}}{2}+\frac{\pi}{{ak}_\mathrm{F}}\right)}-A\left(\frac{\pi}{\mu_\mathrm{h} B}\right)A\left(\frac{T}{T_a}\right)\cos{\left(r_\mathrm{c}+\frac{\delta_\mathrm{F}}{2}-\frac{\pi}{{ak}_\mathrm{F}}\right)}\right], \label{sgmyx22}
\end{equation}
\end{subequations}
\end{widetext}
where $\nu = n_e h/eB$ represents the Landau-level filling fraction, $\lambda_\mathrm{c} \equiv 2\sqrt{2/\pi} [V_0/ (\hbar\omega_c)] [\pi/(ak_\mathrm{F})]{r_\mathrm{c}}^{-1/2}$, $\delta_\mathrm{F} \equiv 2 \cot^{-1}{r_\mathrm{c}}$, and $\mathrm{sgn}(x)$ represents the sign of $x$. We introduced yet another parameter $\mu_\mathrm{h}$ for the damping factor due to the small-angle scattering. We have recently found that experimentally obtained $\delta \sigma_{yx}$ is fairly in good agreement with $\delta \sigma_{yx}$ calculated with $\mu_\mathrm{h} = \mu_{b}/2$ \cite{Endo21HallCO}, indicating that the small-angle scattering diminishes CO amplitudes more severely in the collisional contribution than in the band contribution. Noting that both $\delta \sigma_{xx}^\mathrm{col}$ and $\delta \sigma_{yx}$ are stemming from the collisional contribution, we assume  $\mu_\mathrm{c} = \mu_\mathrm{h} = \mu_{b}/2$ in the present paper.
Again, the oscillatory parts $\delta \sigma_{yx}^{\left(1\right)}$, $\delta \sigma_{yx}^{\left(21\right)}$, and $\delta \sigma_{yx}^{\left(22\right)}$ are the second terms in Eqs.\ (\ref{sgmyx1}), (\ref{sgmyx21}) and (\ref{sgmyx22}).
Collecting the relevant terms, the three independent components of the electrical conductivity tensor are given as
\begin{subequations} \label{sgmall}
\begin{eqnarray}
\sigma_{xx} & = & \sigma_{xx}^\mathrm{sc} + \sigma_{xx}^\mathrm{col}, \label{sgmxx} \\
\sigma_{yy} & = & \sigma_{xx}^\mathrm{sc} + \sigma_{xx}^\mathrm{col} + \sigma_{yy}^\mathrm{band} \label{sgmyy}, \\
\sigma_{yx} & = & -\sigma_{xy} = \sigma_{yx}^\mathrm{sc} + \sigma_{yx}^{\left(1\right)} + \sigma_{yx}^{\left(21\right)} + \sigma_{yx}^{\left(22\right)}, \label{sgmyx}
\end{eqnarray}
\end{subequations}
where $\sigma_{xx}^\mathrm{sc} = \sigma_0 / (1 + \mu^2 B^2)$ and $\sigma_{yx}^\mathrm{sc} = \sigma_0 \mu B / (1 + \mu^2 B^2) \simeq \nu (e^2/h)$ are the diagonal and the off-diagonal components, respectively, of the semiclassical electrical conductivity for an unmodulated 2DEG\@. The corresponding oscillatory parts are $\delta \sigma_{xx} = \delta \sigma_{xx}^\mathrm{col}$, $\delta \sigma_{yy} = \delta \sigma_{xx}^\mathrm{col} + \delta \sigma_{yy}^\mathrm{band}$, and $\delta \sigma_{yx} = \delta \sigma_{yx}^{\left(1\right)} + \delta \sigma_{yx}^{\left(21\right)} + \delta \sigma_{yx}^{\left(22\right)}$.

The resistivity tensor $\hat{\rho}$ is obtained by inverting the electrical conductivity tensor $\hat{\sigma}$, Eq.\ (\ref{rhosgm}). 
Assuming that $|\delta \sigma_{ij}| \ll |\sigma_{ij}|$, the oscillatory parts of $\hat{\rho}$ are given, to the lowest order in $\delta \sigma_{ij}$ \footnote{Or, to be more precise, to the lowest order in the perturbations caused by $V(x)$. The effect of the non-oscillatory background induced by $V(x)$ on the oscillation is the second-order perturbation and thus neglected}, as 
\begin{subequations} \label{drhoall}
\begin{eqnarray}
\delta \rho_{xx}  = & \!\!\!\!\!\!\!\!\!\!\!\!\!\!\!\!\!\!\!\!\!\!\!\!\!\!\!\!\!\!\!\! {\rho_0}^2 (-\delta\sigma_{xx}+\mu^2B^2\delta\sigma_{yy}-2\mu B\delta\sigma_{yx}), \\
\delta \rho_{yy} = & \!\!\!\!\!\!\!\!\!\!\!\!\!\!\!\!\!\!\!\!\!\!\!\!\!\!\!\!\!\!\!\! {\rho_0}^2 (-\delta\sigma_{yy}+\mu^2B^2\delta\sigma_{xx}-2\mu B\delta\sigma_{yx}), \\
\delta \rho_{xy} = & -\delta \rho_{yx} = 
{\rho_0}^2 [\left(1-\mu^2B^2\right)\delta\sigma_{yx}-\mu B\left(\delta\sigma_{xx}+\delta\sigma_{yy}\right)], \nonumber \\
\end{eqnarray}
\end{subequations}
where $\rho_0 \equiv 1/\sigma_0$ is the resistivity at $B = 0$.

In Figs.\ \ref{Gsgm} and \ref{Grho}, we plot $\delta \sigma_{ij}$ and $\delta \rho_{ij}$ calculated for the temperatures $T = 0.02$, 1.5, 4.2, 10 K, employing the parameters for a typical ULSL sample \cite{KoikeMT,KoikeEP2DS20}: $V_0 = 0.26$ meV, $a = 200$ nm, $\mu_\mathrm{b} = 5.3$ m$^2$/(Vs), $\mu_\mathrm{c} = \mu_\mathrm{h} = \mu_\mathrm{b}/2$, 
$n_e = 4.2\times10^{15}$ m$^{-2}$, and $\mu = 72$ m$^2$/(Vs). 
The figures exhibit well-established behaviors of COs in the electrical conductivity and the resistivity \cite{Peeters92,Endo21HallCO}: $\delta \sigma_{yy}$ and $\delta \rho_{xx}$ arising mainly from the band contribution oscillate with minima at the flat-band conditions and with the amplitudes far exceeding the other components showing (nearly) anti-phase oscillations. (Strictly speaking, the phases of $\delta \sigma_{yx}$ and $\delta \rho_{xy}$ slightly depend on $B$. See Ref.\ \onlinecite{Endo21HallCO} for more details.)
Compared with the original theory \cite{Peeters92}, the inclusion of the additional damping factor $A[\pi/(\mu_sB)]$ ($s = \mathrm{b}, \mathrm{c}, \mathrm{h}$) makes the damping of the oscillations with decreasing magnetic field more rapid in $\delta \rho_{ij}$ and suppresses the divergence of the oscillation amplitude at low magnetic field in $\delta \sigma_{ij}$. Furthermore, the heavier damping of the collisional contributions compared to the band contribution,  recently found experimentally \cite{Endo21HallCO} and accounted for by the relation $\mu_\mathrm{c} = \mu_\mathrm{h} = \mu_{b}/2$, enhances the dominance of $\delta \sigma_{yy}$ and $\delta \rho_{xx}$ over the other components. As will be shown below, this enhancement exerts a significant impact on the behaviors of $\delta S_{ij}$.

\subsection{Thermoelectric conductivity}
Now we are ready to calculate the thermoelectric conductivity. By substituting Eqs.\ (\ref{sgmyybandSp}), (\ref{sgmxxcolSp}), and (\ref{sgmyxallSp}) [corresponding to Eqs.\ (\ref{sgmyyband}), (\ref{sgmxxcol}), and (\ref{sgmyxall}) at $T = 0$, respectively, with the suitable $E$-dependence of the quantities involved] into Eq.\ (\ref{epsilondef}), we obtain the band and the collisional contributions for the diagonal components, 
\begin{widetext}
\begin{align}
& \varepsilon_{yy}^\mathrm{band}=-\frac{L_0eT}{E_\mathrm{F}} \frac{\sigma_0 V_0^2}{E_\mathrm{F} \hbar \omega_\mathrm{c} a k_\mathrm{F}}  \times \nonumber \\
& \left\{ \left(p-\frac{1}{2}\right) + \left[\left(p-\frac{1}{2}\right)A\left(\frac{\pi}{\mu_\mathrm{b} B}\right)+\frac{p_\mathrm{b}}{3}\left(\frac{\pi}{\mu_\mathrm{b} B}\right)^2C\left(\frac{\pi}{\mu_\mathrm{b} B}\right)\right]D\left(\frac{T}{T_a}\right)\sin{r_\mathrm{c}}
+A\left(\frac{\pi}{\mu_\mathrm{b} B}\right)C\left(\frac{T}{T_a}\right)\frac{r_\mathrm{c}}{2}\cos{r_\mathrm{c}}\right\} \label{epsyyband}
\end{align}
and
\begin{align}
& \varepsilon_{xx}^\mathrm{col}=-\frac{eL_0T}{E_\mathrm{F}} \frac{3 \sigma_0 V_0^2 a k_\mathrm{F}}{8 \pi^2 E_\mathrm{F} \hbar \omega_\mathrm{c} \mu^2 B^2}  \times \nonumber \\
& \left\{ -\left(p-\frac{1}{2}\right) - \left[-\left(p-\frac{1}{2}\right)A\left(\frac{\pi}{\mu_\mathrm{c} B}\right)+\frac{p_\mathrm{c}}{3}\left(\frac{\pi}{\mu_\mathrm{c} B}\right)^2C\left(\frac{\pi}{\mu_\mathrm{c} B}\right)\right] D\left(\frac{T}{T_a}\right)\sin{r_\mathrm{c}}-A\left(\frac{\pi}{\mu_\mathrm{c} B}\right)C\left(\frac{T}{T_a}\right)\frac{r_\mathrm{c}}{2}\cos{r_\mathrm{c}}\right\} \label{epsxxcol}
\end{align}
\end{widetext}
as well as the three constituents for the off-diagonal component,
\begin{widetext}
\begin{subequations} \label{epsyxAll}
\begin{align}
\varepsilon_{yx}^{(1)} & =  -\frac{L_0eT}{E_\mathrm{F}} \frac{3}{4} \nu \frac{e^2}{h} {\lambda_\mathrm{c}}^2
 \left\{-1+\left[A\left(\frac{\pi}{\mu_\mathrm{h} B}\right)-\frac{2 p_\mathrm{h}}{3}\left(\frac{\pi}{\mu_\mathrm{h} B}\right)^2C\left(\frac{\pi}{\mu_\mathrm{h} B}\right)\right]D\left(\frac{T}{T_a}\right)\sin{\left(r_\mathrm{c}+\delta_\mathrm{F}\right)} \right.  \nonumber \\
 & \left. -A\left(\frac{\pi}{\mu_\mathrm{h} B}\right)C\left(\frac{T}{T_a}\right)r_\mathrm{c}\cos{\delta_\mathrm{F}}\cos{\left(r_\mathrm{c}+\delta_\mathrm{F}\right)} \right\}, \label{epsyx1}
\end{align}
\begin{align}
\varepsilon_{yx}^{(21)} & = -\frac{L_0eT}{E_\mathrm{F}}\frac{1}{2}\nu \frac{e^2}{h}{\lambda_\mathrm{c}}^2 \left\{ \cos{\left(\frac{\delta_\mathrm{F}}{2}+2\frac{\pi}{ak_\mathrm{F}}\right)} + A\left(\frac{\pi}{\mu_\mathrm{h} B}\right)C\left(\frac{T}{T_a}\right)\sin{\left(r_\mathrm{c}+\frac{\delta_\mathrm{F}}{2}-2\frac{\pi}{ak_\mathrm{F}}\right)}  \right. \nonumber \\
& -\frac{ak_\mathrm{F}}{\pi}\sin{\left(\frac{\pi}{ak_\mathrm{F}}\right)} \left[ \frac{1}{2}\sin{\delta_\mathrm{F}}\sin{\left(\frac{\delta_\mathrm{F}}{2}+\frac{\pi}{ak_\mathrm{F}} \right)} + \frac{2p_\mathrm{h}}{3}\left(\frac{\pi}{\mu_\mathrm{h} B}\right)^2C\left(\frac{\pi}{\mu_\mathrm{h} B}\right)D\left(\frac{T}{T_a}\right)\sin{\left(r_\mathrm{c}+\frac{\delta_\mathrm{F}}{2}-\frac{\pi}{ak_\mathrm{F}}\right)} \right. \nonumber \\
& \left. \left. + A\left(\frac{\pi}{\mu_\mathrm{h} B}\right)C\left(\frac{T}{T_a}\right)r_\mathrm{c}\cos^2{\left( \frac{\delta_\mathrm{F}}{2} \right)}\cos{\left(r_\mathrm{c}+\frac{\delta_\mathrm{F}}{2}-\frac{\pi}{ak_\mathrm{F}}\right)} \right] \right\}, \label{epsyx21}
\end{align}
and
\begin{align}
\varepsilon_{yx}^{(22)} & = -\mathrm{sgn}(B)\frac{L_0eT}{E_\mathrm{F}}\frac{1}{2}\frac{e^2}{h}{\lambda_\mathrm{c}}^2 \left( \cos{\left(\frac{\delta_\mathrm{F}}{2}+2\frac{\pi}{ak_\mathrm{F}}\right)} - A\left(\frac{\pi}{\mu_\mathrm{h} B}\right)C\left(\frac{T}{T_a}\right)\sin{\left(r_\mathrm{c}+\frac{\delta_\mathrm{F}}{2}-2\frac{\pi}{ak_\mathrm{F}}\right)}  \right. \nonumber \\
& -\frac{ak_\mathrm{F}}{\pi}\cos{\left(\frac{\pi}{ak_\mathrm{F}}\right)} \left\{ -2\sin{\left(\frac{\delta_\mathrm{F}}{2}+\frac{\pi}{ak_\mathrm{F}} \right)} -\frac{1}{2}\sin{\delta_\mathrm{F}}\cos{\left(\frac{\delta_\mathrm{F}}{2}+\frac{\pi}{ak_\mathrm{F}} \right)} \right. \nonumber \\
& + \left[ 2 A\left(\frac{\pi}{\mu_\mathrm{h} B}\right)-\frac{2p_\mathrm{h}}{3}\left(\frac{\pi}{\mu_\mathrm{h} B}\right)^2C\left(\frac{\pi}{\mu_\mathrm{h} B}\right) \right] D\left(\frac{T}{T_a}\right)\cos{\left(r_\mathrm{c}+\frac{\delta_\mathrm{F}}{2}-\frac{\pi}{ak_\mathrm{F}}\right)}  \nonumber \\
& \left. \left. + A\left(\frac{\pi}{\mu_\mathrm{h} B}\right)C\left(\frac{T}{T_a}\right)r_\mathrm{c}\cos^2{\left( \frac{\delta_\mathrm{F}}{2} \right)}\sin{\left(r_\mathrm{c}+\frac{\delta_\mathrm{F}}{2}-\frac{\pi}{ak_\mathrm{F}}\right)} \right\} \right). \label{epsyx22}
\end{align}
\end{subequations}
\end{widetext}
Here we introduced two new functions,
\begin{widetext}
\begin{equation}
C\left(x\right)\equiv-3\frac{A^\prime\left(x\right)}{x}=\frac{3}{\sinh{x}}\left(\coth{x}-\frac{1}{x}\right)=1-\frac{7}{30}x^2+O\left(x^4\right)  \label{EqC}
\end{equation}
and
\begin{equation}
D\left(x\right)\equiv-3 A^{\prime\prime}\left(x\right)=\frac{3}{\sinh{x}}\left[2\coth{x}-\frac{2x}{\sinh^2{x}}-x\right]=1-\frac{7}{10}x^2+O\left(x^4\right) \label{EqD}
\end{equation}
\end{widetext}
describing the damping, which is plotted in Fig.\ \ref{ACD} along with $A(x)$ in Eq.\ (\ref{EqA}). We assumed, as is usually done \cite{GoldsmidBk10}, power-law dependence, $\mu \propto E^p$ and $\mu_s \propto E^{p_s}$ ($s = \mathrm{b}, \mathrm{c}, \mathrm{h}$), of the mobilities on the energy $E$. The oscillatory parts $\delta \varepsilon_{yy}^\mathrm{band}$, $\delta \varepsilon_{xx}^\mathrm{col}$, $\delta \varepsilon_{yx}^{\left(1\right)}$, $\delta \varepsilon_{yx}^{\left(21\right)}$, and $\delta \varepsilon_{yx}^{\left(22\right)}$ are obtained by collecting the cosine and sine terms containing $r_\mathrm{c}$ in the argument. The three independent components of $\hat{\varepsilon}$ are thus given by
\begin{subequations} \label{epsall}
\begin{eqnarray}
\varepsilon_{xx} & = & \varepsilon_{xx}^\mathrm{sc} + \varepsilon_{xx}^\mathrm{col}, \label{epsxx} \\
\varepsilon_{yy} & = & \varepsilon_{xx}^\mathrm{sc} + \varepsilon_{xx}^\mathrm{col} + \varepsilon_{yy}^\mathrm{band} \label{epsyy}, \\
\varepsilon_{yx} & = & -\varepsilon_{xy} = \varepsilon_{yx}^\mathrm{sc} + \varepsilon_{yx}^{\left(1\right)} + \varepsilon_{yx}^{\left(21\right)} + \varepsilon_{yx}^{\left(22\right)}, \label{epsyx}
\end{eqnarray}
\end{subequations}
where $\varepsilon_{xx}^\mathrm{sc} = -(L_0 e T/E_\mathrm{F})[1+p(1-\mu^2B^2)/(1+\mu^2B^2)] \sigma_{xx}^\mathrm{sc}$ and $\varepsilon_{yx}^\mathrm{sc} = -(L_0 e T/E_\mathrm{F})[1+2p/(1+\mu^2B^2)] \sigma_{yx}^\mathrm{sc} \simeq  -(L_0 e T/E_\mathrm{F}) \nu (e^2/h)$ are the diagonal and the off-diagonal components of the semiclassical thermoelectric conductivity tensors for an unmodulated 2DEG\@. The oscillatory parts are $\delta \varepsilon_{xx} = \delta \varepsilon_{xx}^\mathrm{col}$, $\delta \varepsilon_{yy} = \delta \varepsilon_{xx}^\mathrm{col} + \delta \varepsilon_{yy}^\mathrm{band}$, and $\delta \varepsilon_{yx} = \delta \varepsilon_{yx}^{\left(1\right)} + \delta \varepsilon_{yx}^{\left(21\right)} + \delta \varepsilon_{yx}^{\left(22\right)}$. The oscillation phase of $\delta \varepsilon_{xx}$ and $\delta \varepsilon_{yy}$ are rather complicated, since the relative weight of the terms with $\sin r_\mathrm{c}$ and $\cos r_\mathrm{c}$ in $\delta \varepsilon_{yy}^\mathrm{band}$ and $\delta \varepsilon_{xx}^\mathrm{col}$ varies with both $B$ and $T$. We can find, however, that the oscillations are dominated by the last terms in Eqs.\ (\ref{epsyyband}) and (\ref{epsxxcol}), noting that $r_\mathrm{c}$ generally takes a large value when COs are observed.  Roughly speaking, therefore, $\delta \varepsilon_{yy}^\mathrm{band}$ and $\delta \varepsilon_{xx}^\mathrm{col}$ oscillate with opposite phases and with the amplitude $|\delta \varepsilon_{yy}^\mathrm{band}| \gg |\delta \varepsilon_{xx}^\mathrm{col}|$, similar to the case in $\hat{\sigma}$, leading to antiphase oscillations in $\delta \varepsilon_{xx}$ and $\delta \varepsilon_{yy}$ with $|\delta \varepsilon_{yy}| \gg |\delta \varepsilon_{xx}|$. The phases of $\delta \varepsilon_{xx}$ and $\delta \varepsilon_{yy}$ are approximately $\pi/2$ retarded compared to their counterparts in $\delta \sigma_{xx}$ and $\delta \sigma_{yy}$. The phase of $\delta \varepsilon_{yx}$ is further complicated by the $B$-dependent $\delta_\mathrm{F}$ in the argument of cosine and sine terms. As we pointed out previously \cite{Endo21HallCO}, however, both $\delta_\mathrm{F}$ and $\pi/(ak_\mathrm{F})$ contained in the arguments are expected to be small for typical ULSLs. Noting also that $\delta \varepsilon_{yx}^{\left(22\right)}$ is much smaller than $\delta \varepsilon_{yx}^{\left(1\right)}$ and $\delta \varepsilon_{yx}^{\left(21\right)}$ due to the lack of the factor $\nu$, we find that the dominant terms are the last terms in Eq.\ (\ref{epsyx1}) and in Eq.\ (\ref{epsyx21}), and both roughly oscillate as $\cos r_\mathrm{c}$. Therefore, the phase of $\delta \varepsilon_{yx}$ is close to that of $\delta \varepsilon_{xx}$, as expected from their common origin (collisional contribution).
\begin{figure}[tbh]
\includegraphics[width=8.6cm,clip]{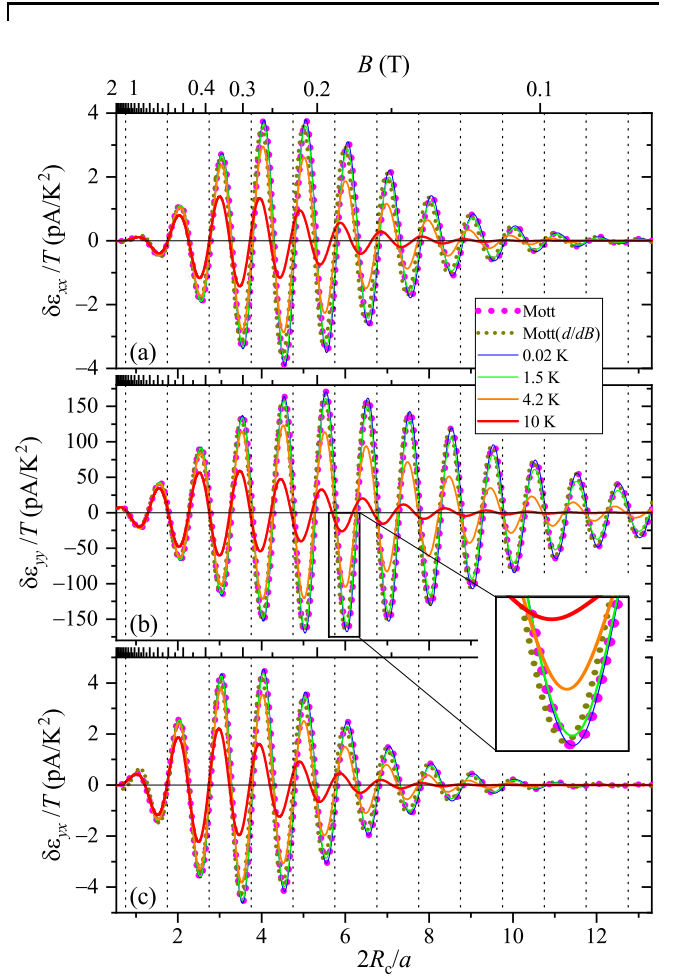}%
\caption{(Color online) Commensurability oscillations in the thermoelectric conductivity divided by the temperature vs.\ $2 R_\mathrm{c} / a$ calculated for various temperatures by picking out the oscillatory parts from Eq.\ (\ref{epsall}) and using the same ULSL parameters as in Fig.\ \ref{Gsgm} and $p = p_\mathrm{b} = p_\mathrm{c} = p_\mathrm{h} = 3/2$. The top axis shows $B$. (a) $\delta \varepsilon_{xx} / T$, (b) $\delta \varepsilon_{yy} / T$, and (c) $\delta \varepsilon_{yx} / T$. $\delta \varepsilon_{ij}^\mathrm{M} / T$ and $\delta \varepsilon_{ij}^\mathrm{MB} / T$ calculated employing the Mott formula [Eq.\ (\ref{MottRel})] and the modified Mott formula [Eq.\ (\ref{MottRelB})] are plotted by thick-dotted (Mott) and thin-dotted [Mott($d/dB$)] lines, respectively. The inset to (b) shows the close up of the part enclosed by the box in the main panel. Vertical dotted lines indicate the locations of the flat-band conditions Eq.\ (\ref{flatband}). \label{Geps}}
\end{figure}

We can draw out low-temperature approximations of $\hat{\varepsilon}$ by the Mott formula Eq.\ (\ref{MottRel}). By performing the energy differentiation on Eqs.\ (\ref{sgmyyband}), (\ref{sgmxxcol}), and (\ref{sgmyxall}) at $T = 0$,
we find that the obtained approximations
$\varepsilon_{yy}^\mathrm{M,band}$, $\varepsilon_{xx}^\mathrm{M,col}$, $\varepsilon_{yx}^\mathrm{M(1)}$, $\varepsilon_{yx}^\mathrm{M(21)}$, $\varepsilon_{yx}^\mathrm{M(22)}$ are given by Eqs. (\ref{epsyyband}),  (\ref{epsxxcol}), (\ref{epsyx1}), (\ref{epsyx21}), and (\ref{epsyx22}) with every occurrence of $C(T/T_a)$ and $D(T/T_a)$ replaced by unity. Therefore, the Mott formulas become a good approximation when $C(T/T_a)$ and $D(T/T_a)$ are close enough to 1, or when $T \ll T_a$ (see Fig.\ \ref{ACD}), which guarantees $\sigma_{ij,{T=0}}\left(E\right)$ to vary slowly within the energy window $|E-E_\mathrm{F}| \lesssim k_\mathrm{B}T$. A general condition for the Mott formula to be applicable is $k_\mathrm{B} T \ll E_\mathrm{F}$, which justifies discarding the higher-order terms in the expansion Eq.\ (\ref{sgmexpnd}). Since $E_\mathrm{F}/(k_\mathrm{B}T_a) = \pi^2 (2R_\mathrm{c}/a) \gg 1$ when COs are observed, $T \ll T_a$ poses a much more stringent constraint. 

An alternative convenient way often employed to estimate the thermoelectric conductivity is to use a modified Mott formula with the energy derivative replaced by the magnetic-field derivative \cite{Fujita10E,Pogosov05}. Noting that the behaviors of COs are governed by the variation of the ratio $2 R_\mathrm{c}/a \propto \sqrt{E}/B$, we replace $d/dE$ by $(-B/2E_\mathrm{F})(d/dB)$ \footnote{Note the difference from the similar formula for the SdH effect, in which $d/dE$ is replaced with $(-B/E_\mathrm{F})(d/dB)$ \cite{Fujita10E} since the oscillations are governed by $\nu \propto E/B$}, yielding an alternative approximation formula, 
\begin{equation}
\varepsilon_{i j}^\mathrm{MB} =L_0eT \frac{B}{2E_\mathrm{F}} \left.\frac{\partial\sigma_{i j,{T=0}}}{\partial B}\right|_{E_\mathrm{F}}. \label{MottRelB}
\end{equation}
An advantage of this formula is that it allows us to estimate $\hat{\varepsilon}$ by numerically differentiating the data of $\hat{\sigma}(B)$ obtained (often by inverting the resistivity $\hat{\rho}(B)$) experimentally from the most basic magnetic-field sweeps using a conventional ULSL sample with fixed $n_e$. By contrast, the acquisition of $d/dE$ requires an additional uniform gate to vary $n_e$.
By carrying out the magnetic-field differentiation, we obtain
\begin{widetext}
\begin{equation}
\varepsilon_{yy}^{\mathrm{MB,band}}=-\frac{L_0eT}{E_F}\frac{\sigma_0 V_0^2}{E_\mathrm{F} \hbar \omega_\mathrm{c} a k_\mathrm{F}} \left\{\frac{1}{2}+\frac{1}{2}\left[A\left(\frac{\pi}{\mu_\mathrm{b} B}\right)-\frac{1}{3}\left(\frac{\pi}{\mu_\mathrm{b} B}\right)^2C\left(\frac{\pi}{\mu_\mathrm{b} B}\right)\right]\sin{r_\mathrm{c}}+A\left(\frac{\pi}{\mu_\mathrm{b} B}\right)\frac{r_\mathrm{c}}{2}\cos{r_\mathrm{c}}\ \right\},  \label{epsyyBandB}
\end{equation}
\begin{equation}
\varepsilon_{xx}^{\mathrm{MB,col}}=-\frac{L_0eT}{E_F}\frac{3 \sigma_0 V_0^2 a k_\mathrm{F}}{8 \pi^2 E_\mathrm{F} \hbar \omega_\mathrm{c} \mu^2 B^2} \left\{\frac{3}{2}-\frac{1}{2}\left[3A\left(\frac{\pi}{\mu_\mathrm{c} B}\right)-\frac{1}{3}\left(\frac{\pi}{\mu_\mathrm{c} B}\right)^2C\left(\frac{\pi}{\mu_\mathrm{c} B}\right)\right]\sin{r_\mathrm{c}}-A\left(\frac{\pi}{\mu_\mathrm{c} B}\right)\frac{r_\mathrm{c}}{2}\cos{r_\mathrm{c}}\right\}, \label{epsxxColB}
\end{equation}
\begin{subequations} \label{epsyxB}
\begin{align}
\varepsilon_{yx}^{\mathrm{MB(1)}}=-\frac{L_0eT}{E_F} \frac{3}{4} \nu\frac{e^2}{h}{\lambda_c}^2\left\{2-\left[2A\left(\frac{\pi}{\mu_\mathrm{h} B}\right)-\frac{1}{3}\left(\frac{\pi}{\mu_\mathrm{h} B}\right)^2C\left(\frac{\pi}{\mu_\mathrm{h} B}\right)\right]\sin{\left(r_\mathrm{c}+\delta_\mathrm{F}\right)} \right. \nonumber \\
\left. -A\left(\frac{\pi}{\mu_\mathrm{h} B}\right)r_\mathrm{c}\cos{\delta_\mathrm{F}}\cos{\left(r_\mathrm{c}+\delta_\mathrm{F}\right)}\right\}, \label{epsyx1B}
\end{align}
\begin{align}
\varepsilon_{yx}^{\mathrm{MB(21)}} & = -\frac{L_0eT}{E_F}\frac{1}{2}\nu\frac{e^2}{h}{\lambda_c}^2\frac{ak_\mathrm{F}}{\pi}\sin{\left(\frac{\pi}{ak_\mathrm{F}}\right)}\left\{-2\cos{\left(\frac{\delta_\mathrm{F}}{2}+\frac{\pi}{ak_\mathrm{F}}\right)}-\frac{1}{2}\sin{\delta_\mathrm{F}}\sin{\left(\frac{\delta_\mathrm{F}}{2}+\frac{\pi}{ak_\mathrm{F}}\right)} \right. \nonumber \\
& \left. -\left[2A\left(\frac{\pi}{\mu_\mathrm{h} B}\right)-\frac{1}{3}\left(\frac{\pi}{\mu_\mathrm{h} B}\right)^2C\left(\frac{\pi}{\mu_\mathrm{h} B}\right)\right]\sin{\left(r_\mathrm{c}+\frac{\delta_\mathrm{F}}{2}-\frac{\pi}{ak_\mathrm{F}}\right)}-A\left(\frac{\pi}{\mu_\mathrm{h} B}\right)r_\mathrm{c}\cos^2{\left( \frac{\delta_\mathrm{F}}{2} \right)}\cos{\left(r_\mathrm{c}+\frac{\delta_\mathrm{F}}{2}-\frac{\pi}{ak_\mathrm{F}}\right)}\right\}, \label{epsyx21B}
\end{align}
and
\begin{align}
\varepsilon_{yx}^{\mathrm{MB(22)}} & =-\mathrm{sign}\left(B\right)\frac{L_0eT}{E_F}\frac{1}{2}\frac{e^2}{h}{\lambda_c}^2\frac{ak_\mathrm{F}}{\pi}\cos{\left(\frac{\pi}{ak_\mathrm{F}}\right)} \left\{-\sin{\left(\frac{\delta_\mathrm{F}}{2}+\frac{\pi}{ak_\mathrm{F}}\right)}+\frac{1}{2}\sin{\delta_\mathrm{F}}\cos{\left(\frac{\delta_\mathrm{F}}{2}+\frac{\pi}{ak_\mathrm{F}}\right)} \right. \nonumber \\
& \left. +\left[A\left(\frac{\pi}{\mu_\mathrm{h} B}\right)-\frac{1}{3}\left(\frac{\pi}{\mu_\mathrm{h} B}\right)^2C\left(\frac{\pi}{\mu_\mathrm{h} B}\right)\right]\cos{\left(r_\mathrm{c}+\frac{\delta_\mathrm{F}}{2}-\frac{\pi}{ak_\mathrm{F}}\right)}-A\left(\frac{\pi}{\mu_\mathrm{h} B}\right)r_\mathrm{c}\cos^2{\left( \frac{\delta_\mathrm{F}}{2} \right)}\sin{\left(r_\mathrm{c}+\frac{\delta_\mathrm{F}}{2}-\frac{\pi}{ak_\mathrm{F}}\right)}\right\}. \label{epsyx22B}
\end{align}
\end{subequations}
\end{widetext}
Note that the non-oscillatory parts in Eqs.\ (\ref{epsyyBandB}), (\ref{epsxxColB}), and (\ref{epsyxB}) make little sense, since the replacement $d/dE \rightarrow -(B/2E_\mathrm{F}) d/dB$ is justifiable only for the COs. It can readily be found that the dominant terms in $\delta \varepsilon_{yy}^\mathrm{MB,band}$, $\delta \varepsilon_{xx}^\mathrm{MB,col}$, $\delta \varepsilon_{yx}^{\mathrm{MB}\left(1\right)}$, $\delta \varepsilon_{yx}^{\mathrm{MB}\left(21\right)}$, and $\delta \varepsilon_{yx}^{\mathrm{MB}\left(22\right)}$ [the last terms in Eqs.\ (\ref{epsyyBandB}), (\ref{epsxxColB}), and (\ref{epsyxB})] coincide with the corresponding terms in $\delta \varepsilon_{yy}^\mathrm{M,band}$, $\delta \varepsilon_{xx}^\mathrm{M,col}$, $\delta \varepsilon_{yx}^{\mathrm{M}\left(1\right)}$, $\delta \varepsilon_{yx}^{\mathrm{M}\left(21\right)}$, and $\delta \varepsilon_{yx}^{\mathrm{M}\left(22\right)}$. For completeness, we mention that the components of the thermoelectric conductivity tensor for the (modified) Mott formula, $\varepsilon_{xx}^\mathrm{M(B)}$, $\varepsilon_{yy}^\mathrm{M(B)}$, and $\varepsilon_{yx}^\mathrm{M(B)}$, are obtained by the same relations as Eq.\ (\ref{epsall}).

\begin{figure}[t]
\includegraphics[width=8.6cm,clip]{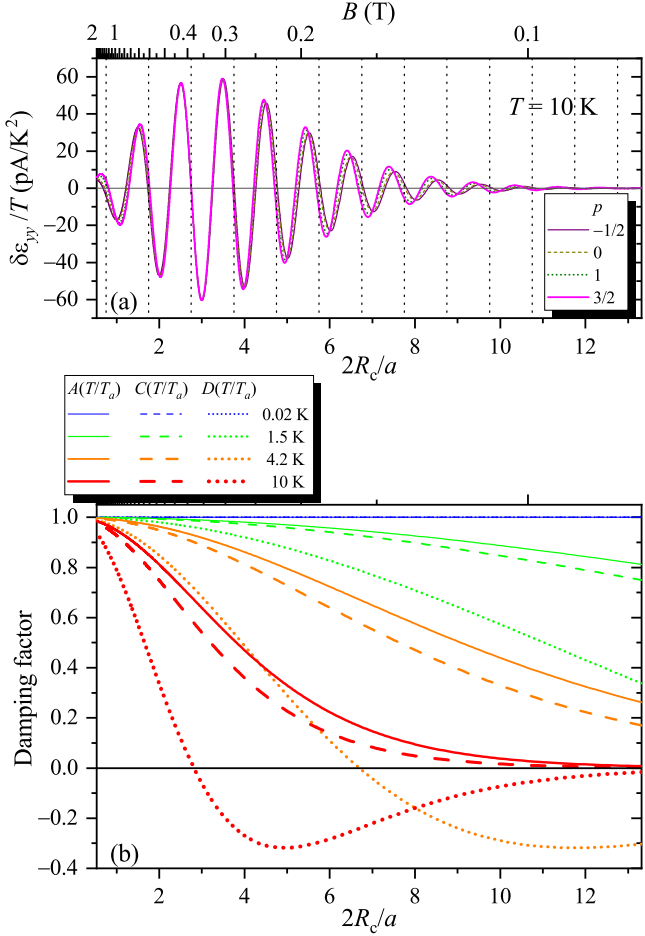}%
\caption{(Color online) (a) $\delta \varepsilon_{yy} / T$ vs.\ $2 R_\mathrm{c} / a$ at $T = 10$ K for various values of the exponent $p (= p_\mathrm{b} = p_\mathrm{c} = p_\mathrm{h})$. The other sample parameters are the same as in Fig.\ \ref{Geps}. The top axis shows $B$. Vertical dotted lines indicate the locations of the flat-band conditions Eq.\ (\ref{flatband}).  (b) Thermal damping factors $A(T/T_a)$, $C(T/T_a)$, and $D(T/T_a)$ vs.\ $2 R_\mathrm{c} / a$ for various temperatures and $T_a = (1/\pi^2)(E_\mathrm{F}/k_\mathrm{B})(a/2R_\mathrm{c})$ with $a =200$ nm and $n_e = 4.2\times10^{15}$ m$^{-2}$ [the same parameters as in (a)], plotted by solid, dashed, and dotted lines, respectively. \label{ACDandphse}}
\end{figure}
In Fig.\ \ref{Geps}, we plot $\delta \varepsilon_{ij} / T$ ($T = 0.02$ K, 1.5 K, 4.2 K, and 10 K), $\delta \varepsilon_{ij}^\mathrm{M} / T$, and $\delta \varepsilon_{ij}^\mathrm{MB} / T$ calculated using the same ULSL parameters employed in Figs.\ \ref{Gsgm} and \ref{Grho} and, in addition, setting the exponents $p = p_\mathrm{b} = p_\mathrm{c} = p_\mathrm{h} = 3/2$, noting that the scatterings are primarily caused by the remote ionized impurities in GaAs/AlGaAs 2DEGs \cite{Davies98B}. As can be seen from Eqs.\ (\ref{MottRel}) and (\ref{MottRelB}), $\delta \varepsilon_{ij}^\mathrm{M} / T$ and $\delta \varepsilon_{ij}^\mathrm{MB} / T$ do not depend on the temperature. 
The figures allow us to confirm the basic characteristics mentioned above, namely, large amplitude oscillations in $\delta \varepsilon_{yy}$ with the phase roughly $\pi/2$ shifted with respect to $\delta \sigma_{yy}$ and much smaller antiphase oscillations in $\delta \varepsilon_{xx}$ and $\delta \varepsilon_{yx}$. Although all the $\delta \varepsilon_{ij}$ are virtually indistinguishable from the corresponding $\delta \varepsilon_{ij}^\mathrm{M}$ at 0.02 K, small but discernible discrepancy becomes apparent already at 1.5 K\@. The oscillation amplitudes exhibit sublinear increase with $T$, thereby enhancing the discrepancy from the Mott formula with $T$. The deviation from the Mott formula varying with the temperature and the magnetic field are visualized by the behavior of $C(T/T_a)$ and $D(T/T_a)$ plotted in Fig.\ \ref{ACDandphse}(b). Deviation of the phase also becomes clear at higher temperatures. The phase shift is caused by the increase in the relative weight of the terms oscillating with $\sim \sin r_\mathrm{c}$. Since the weight depends on the exponents $p$ and $p_s$ ($s = \mathrm{b}, \mathrm{c}, \mathrm{h}$), the phase shift can, in principle, be applied to the experimentally probe the values of the exponents, either to confirm $p = p_s = 3/2$ or to prove otherwise [see Fig.\ \ref{ACDandphse} (a)].
We note in passing that $\delta \varepsilon_{ij}^\mathrm{MB} / T$ provides a fairly good approximation of $\delta \varepsilon_{ij}^\mathrm{M} / T$, albeit with a slight shift in the phase [see the inset to Fig.\ \ref{Geps} (b)].

\subsection{Thermopower}
\begin{figure*}
\includegraphics[width=17cm,clip]{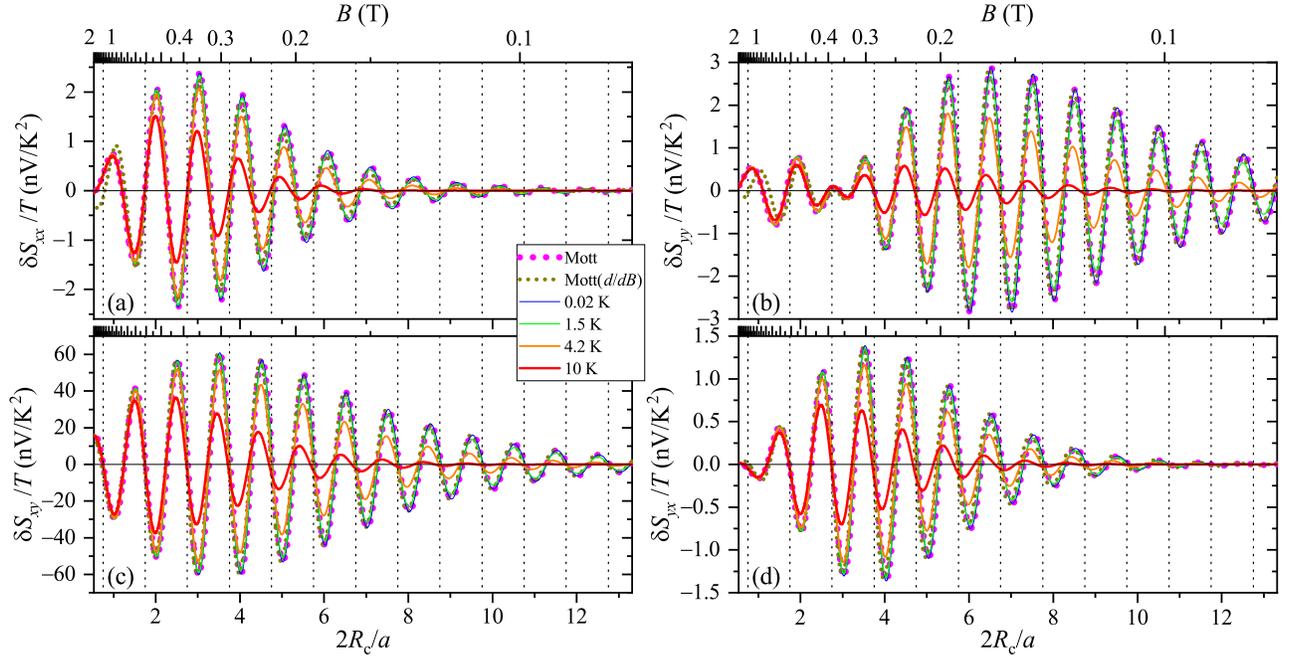}%
\caption{(Color online) Commensurability oscillations in the thermopower divided by the temperature vs.\ $2R_\mathrm{c}/a$ calculated for various temperatures by Eq.\ (\ref{deltaS}) using the same set of parameters as in Fig.\ \ref{Geps}. The top axis shows $B$.  (a) $\delta S_{xx} / T$, (b) $\delta S_{yy} / T$,  (c) $\delta S_{xy} / T$,  and (d)  $\delta S_{yx} / T$. $\delta S_{ij}^\mathrm{M}$ and $\delta S_{ij}^\mathrm{MB}$ calculated by Eq.\ (\ref{deltaS}) with $\delta \varepsilon_{ij}$ replaced by the Mott formulas $\delta \varepsilon_{ij}^\mathrm{M}$ and $\delta \varepsilon_{ij}^\mathrm{MB}$ are also plotted with thick-dotted (Mott) and thin-dotted [Mott($d/dB$)] lines, respectively. Vertical dotted lines indicate the locations of the flat-band conditions Eq.\ (\ref{flatband}). \label{GS}}
\end{figure*}
The thermopower tensor $\hat{S}$ is given by the product of the resistivity tensor and the thermoelectric conductivity tensor, Eq.\ (\ref{Srhoeps}). 
The oscillatory parts of $\hat{S}$ is written, to the lowest order in the perturbations generated by $V(x)$, as
\begin{subequations} \label{deltaS}
\begin{eqnarray}
\delta S_{xx} = & \rho_0 \delta \varepsilon_{xx} + \rho_\mathrm{H} \delta \varepsilon_{yx} + \varepsilon_{xx}^\mathrm{sc} \delta \rho_{xx} + \varepsilon_{yx}^\mathrm{sc} \delta \rho_{xy}, \label{deltaSxx} \\
\delta S_{yy} = & \rho_0 \delta \varepsilon_{yy} + \rho_\mathrm{H} \delta \varepsilon_{yx} + \varepsilon_{xx}^\mathrm{sc} \delta \rho_{yy} + \varepsilon_{yx}^\mathrm{sc} \delta \rho_{xy} , \label{deltaSyy} \\
\delta S_{xy} = & -\rho_0 \delta \varepsilon_{yx} + \rho_\mathrm{H} \delta \varepsilon_{yy} - \varepsilon_{yx}^\mathrm{sc} \delta \rho_{xx}  + \varepsilon_{xx}^\mathrm{sc} \delta \rho_{xy},  \label{deltaSxy} \\
\delta S_{yx} = & \rho_0 \delta \varepsilon_{yx} -\rho_\mathrm{H} \delta \varepsilon_{xx} + \varepsilon_{yx}^\mathrm{sc} \delta \rho_{yy} - \varepsilon_{xx}^\mathrm{sc} \delta \rho_{xy}, \label{deltaSyx}
\end{eqnarray}
\end{subequations}
where $\rho_\mathrm{H} = \rho_{xy}^\mathrm{sc} \equiv B/(en_e)$ is the semiclassical Hall resistivity. In Fig.\ \ref{GS}, we plot $\delta S_{ij}$ calculated by Eq. (\ref{deltaS}) using the same set of sample parameters as in Fig.\ \ref{Geps}. We also plot $\delta S_{ij}^\mathrm{M}$ (Mott) and $\delta S_{ij}^\mathrm{MB}$ [Mott ($d/dB$)] calculated by replacing $\delta \varepsilon_{ij}$ in Eq.\ (\ref{deltaS}) with the Mott formulas $\delta \varepsilon_{ij}^\mathrm{M}$ and $\delta \varepsilon_{ij}^\mathrm{MB}$, respectively.
We can see that the amplitude of $\delta S_{xy}$ exceeds those of the other components by more than an order of magnitude. Accordingly, the relation $S_{xy} = -S_{yx}$, expected for isotropic systems as mentioned earlier, does not hold. We can also find that $\delta S_{xx}$ and $\delta S_{yy}$ exhibit quite dissimilar oscillations. These behaviors are in contradiction to the previous theory \cite{Peeters92}. In Ref.\ \onlinecite{Peeters92}, only $S_{xx}$ and $S_{xy}$ were shown, and $S_{yy}$ were referred to as ``almost identical'' to $S_{xx}$. No mention was made of $S_{yx}$ or its relation to $S_{xy}$. 

We can readily find, by calculating the terms separately, that the first two terms in Eq.\ (\ref{deltaS}) containing $\delta \varepsilon_{ij}$ are much larger than the last two terms with $\delta \rho_{ij}$ \footnote{This can be qualitatively understood by noting that, loosely speaking, $\delta \varepsilon \propto d \delta \sigma /dE$ (Mott formula) while $\delta \rho \propto \delta \sigma$}. Since $|\rho_\mathrm{H}/\rho_0| = \mu |B| \gg 1$ in the magnetic-field range COs are observed, both Eqs.\ (\ref{deltaSxx}) and (\ref{deltaSyy}) are dominated by the identical second term if $|\delta \varepsilon_{xx}|$, $|\delta \varepsilon_{yy}|$, and $|\delta \varepsilon_{yx}|$ are of the same order of magnitude. We surmise that this is the main reason  behind the rather counterintuitive isotropic relation $S_{xx} \simeq S_{yy}$ predicted in Ref.\ \onlinecite{Peeters92}. As we have seen in Fig.\ \ref{Geps}, however, $\delta \varepsilon_{yy}$ possess the amplitude more than an order of magnitude larger than the other components, which inherits the relation $|\delta \sigma_{yy}| \gg |\delta \sigma_{xx}| \sim |\delta \sigma_{yx}|$ in the electrical conductivity (Fig.\ \ref{Gsgm}). As mentioned above, the dominance of the $yy$ component has been intensified by the inclusion of the small-angle-scattering damping factors, which, according to the recent experiment \cite{Endo21HallCO}, diminishes the oscillations of the collisional contribution more effectively compared to those of the band contribution. As a result, while the first term in $\delta S_{xx}$ remains negligibly small, the first term in $\delta S_{yy}$ outweighs the second term at low magnetic fields ($B \lesssim 0.3$ T), leading to the oscillation with the phase roughly opposite to $\delta S_{xx}$. The second term is the dominant term also in both $\delta S_{xy}$ and $\delta S_{yx}$, as was in $\delta S_{xx}$. The second term in $\delta S_{xy}$ consists of the large amplitude oscillations $\delta \varepsilon_{yy}$ multiplied by the large $\rho_\mathrm{H}$, which is responsible for the oscillation amplitude far exceeding those of the other components. $\delta S_{yx}$ oscillates with roughly the same phase and much smaller amplitude. This can readily be understood by recalling the relative phase and amplitude between $\delta \varepsilon_{xx}$ and $\delta \varepsilon_{yy}$ described earlier.

The deviation from the Mott formula with increasing temperature, the phase shift at high temperatures, and the fairly good agreement between $\delta S_{ij}^\mathrm{M}$ and $\delta S_{ij}^\mathrm{MB}$ (albeit with a slight phase shift), all reflects the behaviors of $\delta \varepsilon_{ij}$ described above.

\subsection{Thermal conductivity}
Finally, we calculate the (electronic) thermal conductivity tensors, $\hat{\lambda}$ and $\hat{\kappa}$. Substituting Eqs.\ (\ref{sgmyybandSp}), (\ref{sgmxxcolSp}), and (\ref{sgmyxallSp}) into the definition Eq.\ (\ref{lambdadef}) of $\hat{\lambda}$ and following the procedure similar to that employed in calculating the thermoelectric conductivity, we obtain the band and the collisional contributions for the diagonal components, 
\begin{equation}
\lambda_{yy}^\mathrm{band} = L_0T\frac{\sigma_0 V_0^2}{E_\mathrm{F} \hbar \omega_\mathrm{c} a k_\mathrm{F}}\left[1 + A\left( \frac{\pi}{\mu_\mathrm{b} B} \right) D\left( \frac{T}{T_a} \right) \sin{r_\mathrm{c}} \right], \label{lmdyyband}
\end{equation}
and
\begin{equation}
\lambda_{xx}^\mathrm{col} = L_0T\frac{3 \sigma_0 V_0^2 a k_\mathrm{F}}{8 \pi^2 E_\mathrm{F} \hbar \omega_\mathrm{c} (\mu B)^2}\left[1 - A\left( \frac{\pi}{\mu_\mathrm{c} B} \right) D\left( \frac{T}{T_a} \right) \sin{r_\mathrm{c}} \right], \label{lmdxxcol}
\end{equation}
and the three constituents of the off-diagonal components,
\begin{widetext}
\begin{subequations} \label{lmdyxAll}
\begin{equation}
\lambda_{yx}^{\left(1\right)} = L_0T\frac{3}{2}\nu\frac{e^2}{h}{\lambda_\mathrm{c}}^2\left[1- A\left(\frac{\pi}{\mu_\mathrm{h} B}\right)D\left(\frac{T}{T_a}\right)\sin{\left(r_\mathrm{c}+\delta_\mathrm{F}\right)}\right], \label{lmdyx1}
\end{equation}
\begin{equation}
\lambda_{yx}^{\left(21\right)} = -L_0T\nu\frac{e^2}{h}{\lambda_\mathrm{c}}^2\frac{ak_\mathrm{F}}{\pi}\sin{\left(\frac{\pi}{{ak}_\mathrm{F}}\right)}\left[\cos{\left(\frac{\delta_\mathrm{F}}{2}+\frac{\pi}{{ak}_\mathrm{F}}\right)}+A\left(\frac{\pi}{\mu_\mathrm{h} B}\right)D\left(\frac{T}{T_a}\right)\sin{\left(r_\mathrm{c}+\frac{\delta_\mathrm{F}}{2}-\frac{\pi}{{ak}_\mathrm{F}}\right)}\right], \label{lmdyx21}
\end{equation}
\begin{equation}
\lambda_{yx}^{\left(22\right)} = -L_0T\mathrm{sgn}(B)\frac{e^2}{h}{\lambda_\mathrm{c}}^2\frac{ak_\mathrm{F}}{\pi}\cos{\left(\frac{\pi}{{ak}_\mathrm{F}}\right)}\left[ \sin{\left(\frac{\delta_\mathrm{F}}{2}+\frac{\pi}{{ak}_\mathrm{F}}\right)}-A\left(\frac{\pi}{\mu_\mathrm{h} B}\right)D\left(\frac{T}{T_a}\right)\cos{\left(r_\mathrm{c}+\frac{\delta_\mathrm{F}}{2}-\frac{\pi}{{ak}_\mathrm{F}}\right)}\right]. \label{lmdyx22}
\end{equation}
\end{subequations}
\end{widetext}
The oscillatory parts $\delta \lambda_{yy}^\mathrm{band}$, $\delta \lambda_{xx}^\mathrm{col}$, $\delta \lambda_{yx}^{\left(1\right)}$, $\delta \lambda_{yx}^{\left(21\right)}$, and $\delta \lambda_{yx}^{\left(22\right)}$ are given by the second terms in Eqs.\ (\ref{lmdyyband}), (\ref{lmdxxcol}), (\ref{lmdyx1}), (\ref{lmdyx21}), and (\ref{lmdyx22}), respectively.
The three independent components of $\hat{\lambda}$ are given by
\begin{subequations} \label{lambdaall}
\begin{eqnarray}
\lambda_{xx} & = & \lambda_{xx}^\mathrm{sc} + \lambda_{xx}^\mathrm{col} \label{lmdxx} \\
\lambda_{yy} & = & \lambda_{xx}^\mathrm{sc} + \lambda_{xx}^\mathrm{col} + \lambda_{yy}^\mathrm{band} \label{lmdyy} \\
\lambda_{yx} & = & \lambda_{yx}^\mathrm{sc} + \lambda_{yx}^{\left(1\right)} + \lambda_{yx}^{\left(21\right)} + \lambda_{yx}^{\left(22\right)}, \label{lmdyx}
\end{eqnarray}
\end{subequations}
where $\lambda_{xx}^\mathrm{sc} = L_0T\sigma_{xx}^{sc}$ and $\lambda_{yx}^\mathrm{sc} = L_0T\sigma_{yx}^{sc}$ are the diagonal and off-diagonal compoments of the semiclassical thermal conductivities for an unmodulated 2DEG\@.
The oscillatory parts are $\delta \lambda_{xx} = \delta \lambda_{xx}^\mathrm{col}$,  $\delta \lambda_{yy} = \delta \lambda_{xx}^\mathrm{col} + \delta \lambda_{yy}^\mathrm{band}$, and $\delta \lambda_{yx} = \delta \lambda_{yx}^{\left(1\right)} + \delta \lambda_{yx}^{\left(21\right)} + \delta \lambda_{yx}^{\left(22\right)}$.
By employing the Wiedemann-Franz law Eq.\ (\ref{WFlaw}) instead of Eq.\ (\ref{lambdadef}), we obtain $\lambda_{yy}^\mathrm{W,band}$, $\lambda_{xx}^\mathrm{W,col}$, $\lambda_{yx}^\mathrm{W(1)}$, $\lambda_{yx}^\mathrm{W(21)}$, $\lambda_{yx}^\mathrm{W(22)}$ given by Eqs. (\ref{lmdyyband}),  (\ref{lmdxxcol}), (\ref{lmdyx1}), (\ref{lmdyx21}), and (\ref{lmdyx22}), respectively, with $D(T/T_a)$ replaced by unity. The corresponding three components $\lambda_{ij}^\mathrm{W}$ ($ij = xx, yy, yx$) and their oscillatory parts $\delta \lambda_{ij}^\mathrm{W}$ are obtained with similar relation as in $\lambda_{ij}$ and $\delta \lambda_{ij}$.

Although we referred also to $\hat{\lambda}$ as the ``thermal conductivity'' in this paper (following Ref.\ \onlinecite{Fletcher99}) for lack of a more appropriate nomenclature, thermal conductivity is generally defined as $\hat{\kappa}$ in Eq.\ (\ref{kappalambda}).
Recalling the relation $\varepsilon_{xy} = -\varepsilon_{yx}$ and the symmetry of the diagonal components under magnetic-field reversal, $X_{ii}(-B) = X_{ii}(B)$, we have $^t\! \hat{\varepsilon}(-B) = \hat{\varepsilon}(B)$. 
Further with the help of Eq.\ (\ref{Onsagerpi}), Eq.\ (\ref{kappalambda}) can be rewritten as
\begin{equation}
\hat{\kappa} = \hat{\lambda}-T\hat{\varepsilon}\hat{S}  = \hat{\lambda}-T\hat{\varepsilon}\hat{\rho}\hat{\varepsilon}. \label{kappa}
\end{equation}
The oscillatory parts are
\begin{subequations} \label{deltakappa}
\begin{equation}
\delta \kappa_{xx} = \delta \lambda_{xx} - T (\varepsilon_{xx}^\mathrm{sc} \delta S_{xx} - \varepsilon_{yx}^\mathrm{sc} \delta S_{yx} + S_{xx}^\mathrm{sc} \delta \varepsilon_{xx} - S_{yx}^\mathrm{sc} \delta \varepsilon_{yx}),
\end{equation}
\begin{equation}
\delta \kappa_{yy} = \delta \lambda_{yy} - T (\varepsilon_{xx}^\mathrm{sc} \delta S_{yy} + \varepsilon_{yx}^\mathrm{sc} \delta S_{xy} + S_{xx}^\mathrm{sc} \delta \varepsilon_{yy} - S_{yx}^\mathrm{sc} \delta \varepsilon_{yx}),
\end{equation}
\begin{equation}
\delta \kappa_{yx} = \delta \lambda_{yx} - T (\varepsilon_{xx}^\mathrm{sc} \delta S_{yx} + \varepsilon_{yx}^\mathrm{sc} \delta S_{xx} + S_{yx}^\mathrm{sc} \delta \varepsilon_{yy}  + S_{xx}^\mathrm{sc} \delta \varepsilon_{yx}).
\end{equation}
\end{subequations}
%
%
\begin{figure}[t]
\includegraphics[width=8.6cm,clip]{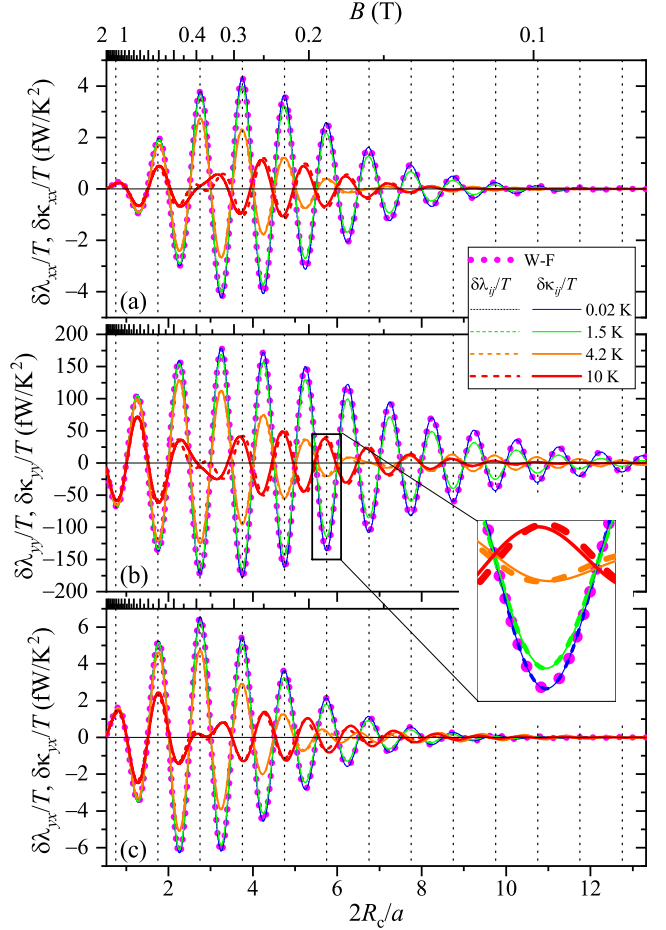}%
\caption{(Color online) Commensurability oscillations in the (electronic) thermal conductivity divided by the temperature vs.\ $2R_\mathrm{c} / a$ calculated for various temperatures. The top axis shows $B$. (a) $\delta \lambda_{xx} / T$ and $\delta \kappa_{xx} / T$, (b) $\delta \lambda_{yx} / T$ and $\delta \kappa_{yx} / T$, (c) $\delta \lambda_{yy} / T$ and $\delta \kappa_{yy} / T$.  $\delta \lambda_{ij} / T$ and $\delta \kappa_{ij} / T$ are plotted by dashed and solid lines, respectively. In addition, $\delta \lambda_{ij}^\mathrm{W} / T$ calculated by applying the Wiedemann-Franz law is plotted by the thick dotted-line (W-F). The inset to (b) displays magnified view of the part enclosed by the rectangle in the main panel. \label{Glmdkappa}}
\end{figure}

In Fig.\ \ref{Glmdkappa}, we plot $\delta \lambda_{ij} / T$ and  $\delta \kappa_{ij} / T$ calculated for various temperatures using the same set of sample parameters as has been used so far. Again, the CO amplitude is overwhelmingly larger in the $yy$ component than in the other components, taking over similar relations in $\hat{\sigma}$. We also plot $\delta \lambda_{ij}^\mathrm{W} / T$, which do not depend on the temperature. We find that $\delta \lambda_{ij} / T$ and  $\delta \kappa_{ij} / T$ are almost indistinguishable at low temperatures, but the latter exhibits a slight phase shift at high temperatures [see the inset to Fig.\ \ref{Glmdkappa}(b)]. Sign reversal takes place for low magnetic fields at high temperatures (4.2 K and 10 K), which is attributable to the sign reversal of $D(T/T_a)$ [see Fig.\ \ref{ACDandphse}(b)]. Deviation of $\delta \lambda_{ij} / T$ and  $\delta \kappa_{ij} / T$ from the Wiedemann-Franz law is discernible even at 1.5 K, resulting from the deviation of $D(T/T_a)$ from unity (see Fig.\ \ref{ACD}).

\section{Discussion}
\begin{figure}[t]
\includegraphics[width=8.6cm,clip]{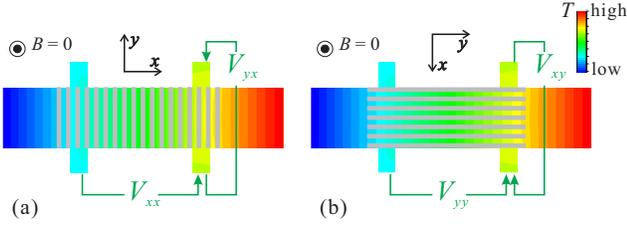}%
\caption{(Color online) Schematic drawings depicting the experimental setup designed to measure the four components of the thermopower tensor $\hat{S}$ in a ULSL\@. The principal axis ($x$ axis) of the ULSL is aligned along and across the Hall bar in (a) and (b), respectively. Temperature gradient $-\boldsymbol{\nabla} T$ is introduced by heating the right end of the Hall bar devices. At $B = 0$, $-\boldsymbol{\nabla} T$ is aligned essentially parallel to the length of the Hall bar, and thus the resulting thermoelectric voltages $V_{xx}$, $V_{yx}$, $V_{yy}$, and $V_{xy}$ represent $S_{xx}$, $S_{yx}$, $S_{yy}$, and $S_{xy}$, respectively, after correcting for the geometrical factors of the devices. \label{SampleB0}}
\end{figure}
\begin{figure}[t]
\includegraphics[width=8.6cm,clip]{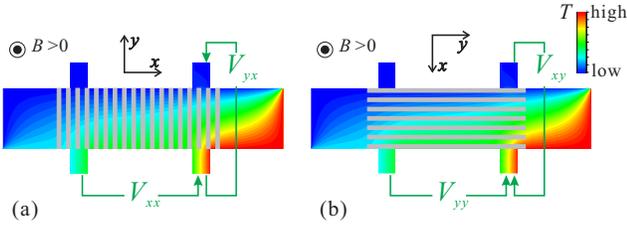}%
\caption{(Color online) Drawings similar to Fig.\ \ref{SampleB0} but with a finite magnetic field $B$ applied perpendicular to the 2DEG plane. The magnetic field deflects $-\boldsymbol{\nabla} T$ away from the length of the Hall bar roughly by the Hall angle $\theta_\mathrm{H}$.  Owing to the tilting, $V_{xx}$, $V_{yx}$, $V_{yy}$, and $V_{xy}$ are not simply related to $S_{xx}$, $S_{yx}$, $S_{yy}$, and $S_{xy}$; $V_{xx}$, for example, contains contributions from both $S_{xx}$ and $S_{xy}$. \label{SampleB05}}
\end{figure}

Here we discuss the implication of the present results on the interpretation of experiments. 
First we recall the practical difficulties one encounters in the measurement involving the temperature gradient in a magnetic field for a device fabricated from a high-mobility 2DEG. Measurement of the thermopower or thermal conductivity using a Hall-bar geometry requires the temperature gradient along the length of the Hall bar. In the absence of the magnetic field, this can readily be achieved by heating one end of the Hall bar (see Fig.\ \ref{SampleB0}). The situation is drastically altered, however, by the application of the magnetic field. When the 2DEG is mostly decoupled from the lattice system, as assumed in the present study, the thermal current is mainly carried by the electrons and thus can readily be bent by the magnetic field. Accordingly, the temperature gradient  $-\boldsymbol{\nabla} T$ generated by heating one end is tilted away from the longitudinal direction of the Hall bar roughly by the Hall angle $\theta_\mathrm{H} = \arctan{(\sigma_{yx} / \sigma_{xx})} $ (see Fig.\ \ref{SampleB05}) \cite{Endo19}. In a high-mobility 2DEG, the relation $|\sigma_{yx}| \gg \sigma_{xx}$ holds in the magnetic field range COs are observed, leading to a high  $|\theta_\mathrm{H}|$ approaching $\pi/2$. Therefore,  $-\boldsymbol{\nabla} T$ is roughly redirected to the transverse direction across the Hall bar, with some position-dependent variation in the direction. 

At $B = 0$, we can employ the pairs of voltage probes located along $V_{xx}$ ($V_{yy}$) or across $V_{yx}$ ($V_{xy}$) the Hall bar to deduce the components of the thermopower tensor $S_{xx}$ ($S_{yy}$) or $S_{yx}$ ($S_{yx}$), respectively, employing a ULSL device with the $x$ axis aligned along (across) the length of the Hall bar as depicted in Fig.\ \ref{SampleB0} (a) [Fig.\ \ref{SampleB0} (b)], and making use of the relation $\boldsymbol{E} = \hat{S} \boldsymbol{\nabla} T$ [Eq.\ (\ref{EjT}) with $\boldsymbol{j} = 0$].
In the magnetic field, however, these correspondences are lost due to the tilt of $-\boldsymbol{\nabla} T$ (Fig.\ \ref{SampleB05}). To the thermovoltage $V_{xx}$ shown in Fig.\ \ref{SampleB05} (a), for instance, not only $S_{xx}$ but also $S_{xy}$ makes a contribution due to the $y$ component $-\nabla_y T$ of $-\boldsymbol{\nabla} T$ introduced by the tilting. Moreover, noting that $|\delta S_{xy}|$ is more than an order of magnitude larger than $|\delta S_{xx}|$ (see Fig.\ \ref{GS}) and that $|\nabla_y T|$ is larger than the $x$ component $|\nabla_x T|$ owing to the high Hall angle, we can conclude that the oscillatory part $\delta V_{xx}$ mostly reflects the behavior of $\delta S_{xy}$ with minor modification attributable to $\delta S_{xx}$. As can be seen in Fig.\ \ref{SampleB05} (b), $\delta V_{xy}$ also picks up $\delta S_{xy}$ and $\delta S_{xx}$, with the relative weight of the latter enhanced since  $|\nabla_x T|$ outweigh  $|\nabla_y T|$ in this setup. In both cases, where the pair of voltage probes are aligned along the $x$ direction, the COs in the thermoelectric voltage are expected to mainly results from $\delta S_{xy}$ owing to its overwhelmingly large oscillation amplitude. On the other hand, thermoelectric voltages measured with the pairs along the $y$ direction, $V_{yx}$ and $V_{yy}$, are composed of contributions from $\delta S_{yx}$ and $\delta S_{yy}$. Even in these cases, however, it is possible for the contribution of $\delta S_{xy}$ to mix into the measured thermoelectric voltage, caused by small misalignment of the locations of the voltage arms or of the orientation of the ULSL, which are difficult to completely eliminate in the actual experiments. The dependence of $-\boldsymbol{\nabla} T$ on the location, as well as its variation with $B$ \cite{Endo19}, further complicates the interpretation of experimentally obtained $V_{ij}$ \cite{KoikeMT,KoikeEP2DS20}.

In Ref.\ \onlinecite{Taboryski95}, the thermoelectric voltage $V_{xx}$ was measured on a ULSL in the configuration shown in Fig.\ \ref{SampleB05} (a). The authors assumed the temperature gradient $-\boldsymbol{\nabla} T$ along the length of the Hall bar [as in Fig.\ \ref{SampleB0} (a)], and attributed the COs observed in $V_{xx}$ to $\delta S_{xx}$. The amplitude and the phase of the COs were found to be roughly in agreement with $\delta S_{xx}$ in the previous theory \cite{Peeters92}. In hindsight, however, the agreement in the amplitude seems to have been caused partly by incorrect analysis in deducing the amplitude $V_0$ of the potential modulation. \footnote{Although lithographically defined period of the grating gate was 400 nm, the period of COs in $R_{xx}$ unambiguously indicates that the period responsible for the COs was $a = 200$ nm, which reveals that the second harmonic content was dominant in the strain-induced modulation potential in this sample. \cite{Larkin97,Endo05HH} Accordingly, $V_0$ determined from the position $B_c$ of the maximum of the low-field positive magnetoresistance should also become half the value quoted in the paper ($V_0 = 0.26$ meV), which, in turn, makes the amplitude of $\delta S_{xx}$ in Ref.\ \onlinecite{Peeters92} smaller} With the correct value of $V_0$, theoretical amplitude becomes smaller than the experimentally-observed amplitude. Moreover, while Ref.\ \onlinecite{Peeters92} predicted the amplitude of the COs in the thermopower to \textit{increase} with decreasing magnetic field, the amplitude of the experimentally observed COs in $V_{xx}$ \textit{decreased} with decreasing magnetic field in contradiction to the theory. This clearly indicates that additional damping factors, such as $A[\pi/(\mu_s B)]$ ($s = \mathrm{b}, \mathrm{c}, \mathrm{h}$) in the present study, are essential for the theory to account for the experimentally observed COs in the thermopower. As discussed above, we consider COs observed in this experiment are also mainly attributable to $\delta S_{xy}$ due to the tilting of $-\boldsymbol{\nabla} T$. In fact, we found that Eq.\ (\ref{deltaSxy}) with the parameters for the sample in Ref.\ \onlinecite{Taboryski95} \footnote{We used the values $V_0 = 0.13$ meV, $a = 200$ nm, $n_e = 2.9\times10^{15}$ m$^{-2}$, and $\mu = 180$ m$^2$/(Vs). We also note that the transverse ($y$) component of $-\boldsymbol{\nabla} T$ changes sign by reversing the magnetic field. Therefore the phase of $\delta S_{xy}$ changes by 180$^\circ$ depending on the direction of the magnetic field, which was not specified in the paper. Here we assumed the direction to be that resulting in the phase close to the experimentally-observed phase of COs in the thermopower} roughly reproduces the amplitude and the phase of the COs presented in Fig.\ 4 of Ref.\ \onlinecite{Taboryski95} including the magnetic-field dependence of the amplitude, assuming the damping parameters to be typical values for GaAs/AlGaAs 2DEGs, $\mu_\mathrm{b} = 2 \mu_\mathrm{c} = 2 \mu_\mathrm{h} = 6.0$ m$^2$/(Vs).

\section{Conclusions}
We have deduced asymptotic analytic expressions for the COs in $\hat{\varepsilon}$ and $\hat{\lambda}$ with Eqs.\ (\ref{epsilondef}) and (\ref{lambdadef}), respectively, 
from those of the electrical conductivity tensor $\hat{\sigma}$ in the literature \cite{Peeters92,Endo21HallCO} slightly modified by introducing an additional damping factor $A[\pi/(\mu_\mathrm{s}B)]$ ($\mathrm{s} = \mathrm{b}, \mathrm{c}, \mathrm{h}$) 
[Eq.\ (\ref{sgmall})]. 
Employing $\hat{\varepsilon}$  and $\hat{\lambda}$ thus obtained, we have further calculated COs in the thermopower tensor $\hat{S}$ and the thermal conductivity tensor $\hat{\kappa}$ with Eqs.\ (\ref{Srhoeps}) and (\ref{kappalambda}), respectively.
The resulting analytic formulas for $\varepsilon_{ij}$, $\lambda_{ij}$, $S_{ij}$, and $\kappa_{ij}$, namely, Eqs.\ (\ref{epsall}),  (\ref{lambdaall}),  (\ref{deltaS}), and  (\ref{deltakappa}), respectively, allow us to easily find out how the amplitude and the phase of COs in these transport coefficients vary with the temperature, the magnetic field, and the parameters of the ULSL samples.
The analytic expressions also reveal that the temperature range for the Mott formula and the Wiedemann-Franz law to be applicable is limited to $T \ll T_a = \hbar \omega_\mathrm{c} a k_\mathrm{F} / (4\pi^2 k_\mathrm{B})$. For typical ULSL samples, this poses more stringent conditions compared to the general condition $T \ll E_\mathrm{F} / k_\mathrm{B}$. 
We have also shown that an empirical variant of the ``Mott formula'' obtained by replacing $d/dE$ with $(-B/2E_\mathrm{F})(d/dB)$ makes a fairly good approximation to the genuine Mott formula.

For the COs in the thermopower tensor, we have found marked anisotropy between the Seebeck components ($\delta S_{xx} \ne \delta S_{yy}$), in contradiction to the previous theory \cite{Peeters92}. We have also found that  the amplitude of COs in one of the Nernst components ($S_{xy}$) far exceeds those in the other components ($|\delta S_{xy}| \gg |\delta S_{xx}|, |\delta S_{yy}|, |\delta S_{yx}|$). The origin of these properties can be traced back to the damping factor $A[\pi/(\mu_\mathrm{s}B)]$, which accounts for the small-angle nature of the scatterings in a device fabricated from a GaAs/AlGaAs 2DEG wafer and have proven to be indispensable for achieving good agreement between the theoretical and the experimentally obtained COs in the $\rho_{xx}$ \cite{Endo00e} and $\rho_{yx}$ \cite{Endo21HallCO}.
Our recent experimental finding of the extremely small CO in $\rho_{xy}$ \cite{Endo21HallCO} suggests much heavier damping of the collisional contribution compared to the band contribution (quantified by $\mu_\mathrm{c} = \mu_\mathrm{h} = \mu_\mathrm{b}/2$), which reinforces the dominance of $|\delta X_{yy}|$ over $|\delta X_{xx}|$ and  $|\delta X_{yx}|$ for $\hat{X} = \hat{\sigma}$, $\hat{\varepsilon}$, and $\hat{\lambda}$. Due to the relation $|\delta \varepsilon_{yy}| \gg |\delta \varepsilon_{xx}|, |\delta \varepsilon_{yx}|$, the relative weight of the term $\rho_\mathrm{H} \delta\varepsilon_{yx}$, which both $\delta S_{xx}$ and $\delta S_{yy}$ have in common [see Eqs.\ (\ref{deltaSxx}) and (\ref{deltaSyy})], is heavily diminished, leading to the prominent anisotropy. Noting that $\rho_\mathrm{H} \gg \rho_0$, we can readily see that $\rho_\mathrm{H} \delta \varepsilon_{yy}$ possesses by far the largest amplitude among all the terms in Eq.\ (\ref{deltaS}), letting $\delta S_{xy}$, the only component containing this term, far surpass $\delta S_{xx}$, $\delta S_{yy}$, and $\delta S_{yx}$ in the amplitude.  We stress that COs in ULSLs provide a prototypical example of anisotropic systems lacking simple relations between the two Nernst coefficients, $S_{xy}$ and $S_{yx}$. Finally, we have discussed the difficulties that obstruct the attempts to experimentally determine $\delta S_{ij}$'s and the possibility that observed COs are dominated by $\delta S_{xy}$ regardless of the configurations of the voltage probes or the orientation of the ULSL\@.

\begin{acknowledgments}
This work was supported by JSPS KAKENHI Grant Numbers 20K03817, 22H01805, and 19H00652.
\end{acknowledgments}

\appendix*
\section{Electrical Conductivity at $T = 0$ as a Function of the Energy}
In this Appendix, we present, for completeness, the explicit expressions for the electrical conductivity at $T = 0$ as a function of the energy (spectral conductivity), $\sigma_{ij,{T=0}}\left(E\right)$, used to calculate $\varepsilon_{ij}$ and $\lambda_{ij}$ via Eqs.\ (\ref{epsilondef}) and (\ref{lambdadef}), respectively. The band and collisional components for the diagonal components, corresponding to Eqs.\ (\ref{sgmyyband}) and (\ref{sgmxxcol}) at $T=0$, respectively, are
\begin{widetext}
\begin{equation}
\sigma_{yy,{T=0}}^\mathrm{band}(E) = \frac{2}{\pi} \frac{e^2}{h} \frac{{V_0}^2}{(\hbar \omega_\mathrm{c})^2} \tilde{\mu} |B|  \alpha E^{p-1/2} \left[1 + A\left( \frac{\pi}{\tilde{\mu}_\mathrm{b}B E^{p_\mathrm{b}}} \right)  \sin{\left( \frac{8\alpha \sqrt{E}}{\hbar \omega_\mathrm{c}}  \right)} \right] \label{sgmyybandSp}
\end{equation}
and
\begin{equation}
\sigma_{xx,{T=0}}^\mathrm{col}(E) = \frac{3 }{4 \pi} \frac{e^2}{h}  \frac{{V_0}^2}{(\hbar \omega_\mathrm{c})^2}  \frac{E^{-(p-1/2)}}{ \tilde{\mu} |B| \alpha }\left[1 - A\left( \frac{\pi}{\tilde{\mu}_\mathrm{c} B E^{p_\mathrm{c}}} \right) \sin{\left( \frac{8\alpha \sqrt{E}}{\hbar \omega_\mathrm{c}}\right)} \right], \label{sgmxxcolSp}
\end{equation}
while the three constituents of the off-diagonal components corresponding to Eqs.\ (\ref{sgmyx1}), (\ref{sgmyx21}), and (\ref{sgmyx22}) are given by
\begin{subequations} \label{sgmyxallSp}
\begin{equation}
\sigma_{yx,{T=0}}^{\left(1\right)}(E) = \mathrm{sgn}(B) \frac{3}{\pi} \frac{e^2}{h} \frac{{V_0}^2}{(\hbar \omega_\mathrm{c})^2}  \frac{\alpha}{\sqrt{E}} \left\{1- A\left(\frac{\pi}{\tilde{\mu}_\mathrm{h}B E^{p_\mathrm{h}}} \right)\sin{\left[ \frac{8\alpha \sqrt{E}}{\hbar \omega_\mathrm{c}}+\delta(E)\right]}\right\}, \label{sgmyx1Sp}
\end{equation}
\begin{equation}
\sigma_{yx,{T=0}}^{\left(21\right)}(E) = -\mathrm{sgn}(B) \frac{2}{\pi} \frac{e^2}{h}  \frac{{V_0}^2}{(\hbar \omega_\mathrm{c})^2}  \sin{\left( \frac{\alpha}{\sqrt{E}} \right) }\left\{ \cos{\left[ \frac{\delta(E)}{2}+\frac{\alpha}{\sqrt{E}}  \right] }+A\left(\frac{\pi}{\tilde{\mu}_\mathrm{h}B E^{p_\mathrm{h}}}\right)\sin{ \left[ \frac{8\alpha \sqrt{E}}{\hbar \omega_\mathrm{c}} +\frac{\delta(E)}{2}-\frac{\alpha}{\sqrt{E}}  \right] }\right\}, \label{sgmyx21Sp}
\end{equation}
and
\begin{equation}
\sigma_{yx,{T=0}}^{\left(22\right)}(E) = -\mathrm{sgn}(B) \frac{1}{\pi} \frac{e^2}{h} \frac{{V_0}^2}{\hbar \omega_\mathrm{c} E}\cos{\left( \frac{\alpha}{\sqrt{E}}  \right) }\left\{ \sin{\left[\frac{\delta(E)}{2}+\frac{\alpha}{\sqrt{E}}  \right]}-A\left(\frac{\pi}{\tilde{\mu}_\mathrm{h}B E^{p_\mathrm{h}}}\right)\cos{\left[ \frac{8\alpha \sqrt{E}}{\hbar \omega_\mathrm{c}}+\frac{\delta(E)}{2}-\frac{\alpha}{\sqrt{E}}  \right]}\right\}, \label{sgmyx22Sp}
\end{equation}
\end{subequations}
\end{widetext}
where $\alpha \equiv \pi \hbar / (a \sqrt{2 m^*})$, $\delta(E) \equiv 2 \cot^{-1} (8\alpha \sqrt{E} / \hbar\omega_\mathrm{c})$, and $\tilde{\mu}$ and $\tilde{\mu}_s$ ($s = \mathrm{b}, \mathrm{c}, \mathrm{h}$) are defined by the relations $\mu = \tilde{\mu} E^p$, $\mu_s = \tilde{\mu}_s E^{p_s}$.

\bibliography{ourpps,thermo,lsls,twodeg,TextBooks,noteAnalTPCO}

\end{document}